\documentclass[fleqn,usenatbib]{mnras}
\usepackage{newtxtext,newtxmath}
\usepackage[T1]{fontenc}

\DeclareRobustCommand{\VAN}[3]{#2}
\let\VANthebibliography\thebibliography
\def\thebibliography{\DeclareRobustCommand{\VAN}[3]{##3}\VANthebibliography}

\usepackage{graphicx}	
\usepackage{amsmath}	
\usepackage{multirow}
\usepackage{soul}
\usepackage[shortcuts]{extdash}
\usepackage{orcidlink}
\usepackage[normalem]{ulem}
\usepackage{eso-pic}

\newcommand{\zobs}{z_{{\rm obs},i}}
\newcommand{\KP}{\citet{DES-SN5YR}} 
\newcommand{\KPp}{\citep{DES-SN5YR}} 
\newcommand{\om}{\Omega_{\rm m}}
\newcommand{\bao}{BAO-$\theta_*$~}
\newcommand{\baoperp}{BAO-$\theta_{*\perp}$~}
\newcommand{\cmb}{CMB-$R$~}

\newcommand{\dl}{D_{\rm L}}
\newcommand{\mate}{Q_H}
\newcommand\scalemath[2]{\scalebox{#1}{\mbox{\ensuremath{\displaystyle #2}}}}

\title[Investigating Beyond-$\Lambda$CDM]{The Dark Energy Survey Supernova Program: Investigating Beyond-$\Lambda$CDM}

\author[R. Camilleri et al.]{
\parbox{\textwidth}{
R.~Camilleri$^{\orcidlink{0000-0002-7436-3950}}$,$^{1}$\thanks{E-mail: uqrcamil@uq.edu.au (RC)}
T.~M.~Davis$^{\orcidlink{0000-0002-4213-8783}}$,$^{1}$
M.~Vincenzi$^{\orcidlink{0000-0001-8788-1688}}$,$^{2}$
P.~Shah$^{\orcidlink{0000-0002-8000-6642}}$,$^{3}$
J.~Frieman$^{\orcidlink{0000-0003-4079-3263}}$,$^{4,5}$
R.~Kessler$^{\orcidlink{0000-0003-3221-0419}}$,$^{6}$
P.~Armstrong$^{\orcidlink{0000-0003-1997-3649}}$,$^{7}$
D.~Brout,$^{8}$
A.~Carr$^{\orcidlink{0000-0003-4074-5659}}$,$^{1}$
R.~Chen,$^{2}$
L.~Galbany,$^{9}$
K.~Glazebrook,$^{10}$
S.~R.~Hinton$^{\orcidlink{0000-0003-2071-9349}}$,$^{1}$
J.~Lee$^{\orcidlink{0000-0001-6633-9793}}$,$^{11}$
C.~Lidman$^{\orcidlink{0000-0003-1731-0497}}$,$^{12,7}$
A.~M\"oller,$^{10}$
B.~Popovic$^{\orcidlink{0000-0002-8012-6978}}$,$^{2}$
H.~Qu,$^{11}$
M.~Sako,$^{11}$
D.~Scolnic,$^{2}$
M.~Smith$^{\orcidlink{0000-0002-3321-1432}}$,$^{13}$
M.~Sullivan$^{\orcidlink{0000-0001-9053-4820}}$,$^{13}$
B.~O.~S\'anchez$^{\orcidlink{0000-0002-8687-0669}}$,$^{2,14}$
G.~Taylor$^{\orcidlink{0000-0001-5756-3259}}$,$^{7}$
M.~Toy$^{\orcidlink{0000-0001-6882-0230}}$,$^{13}$
P.~Wiseman,$^{13}$
T.~M.~C.~Abbott,$^{15}$
M.~Aguena,$^{16}$
S.~Allam,$^{4}$
O.~Alves,$^{17}$
J.~Annis,$^{4}$
S.~Avila,$^{18}$
D.~Bacon,$^{19}$
E.~Bertin,$^{20,21}$
S.~Bocquet,$^{22}$
D.~Brooks,$^{3}$
D.~L.~Burke,$^{23,24}$
A.~Carnero~Rosell,$^{25,16}$
J.~Carretero,$^{18}$
F.~J.~Castander,$^{26,9}$
L.~N.~da Costa,$^{16}$
M.~E.~S.~Pereira,$^{27}$
S.~Desai,$^{28}$
H.~T.~Diehl,$^{4}$
P.~Doel,$^{3}$
C.~Doux,$^{11,14}$
S.~Everett,$^{29}$
I.~Ferrero,$^{30}$
B.~Flaugher,$^{4}$
P.~Fosalba,$^{9}$
J.~Garc\'ia-Bellido,$^{31}$
M.~Gatti,$^{11}$
E.~Gaztanaga,$^{26,19,9}$
G.~Giannini,$^{18,5}$
D.~Gruen,$^{22}$
D.~L.~Hollowood,$^{32}$
K.~Honscheid,$^{33,34}$
D.~J.~James,$^{8}$
K.~Kuehn,$^{35,36}$
O.~Lahav,$^{3}$
S.~Lee,$^{29}$
G.~F.~Lewis,$^{37}$
J.~L.~Marshall,$^{38}$
J. Mena-Fern{\'a}ndez,$^{39}$
R.~Miquel,$^{40,18}$
J.~Muir,$^{41}$
J.~Myles,$^{42}$
R.~L.~C.~Ogando,$^{43}$
A.~Pieres,$^{16}$
A.~A.~Plazas~Malag\'on,$^{23,24}$
A.~Porredon,$^{44}$
M.~Rodriguez-Monroy,$^{31}$
E.~Sanchez,$^{45}$
D.~Sanchez Cid,$^{45}$
M.~Schubnell,$^{17}$
I.~Sevilla-Noarbe,$^{45}$
E.~Suchyta,$^{46}$
M.~E.~C.~Swanson,$^{47}$
G.~Tarle,$^{17}$
A.~R.~Walker,$^{15}$
and N.~Weaverdyck$^{48,49}$ (DES Collaboration)
}
\vspace{0.4cm}
\\
\textit{Affiliations are listed at the end of the paper}
}

\date{Accepted XXX. Received YYY; in original form ZZZ}

\AddToShipoutPictureBG*{%
  \AtPageUpperLeft{%
    \hspace{0.75\paperwidth}%
    \raisebox{-1.5\baselineskip}{%
      \makebox[0pt][l]{\textnormal{DES-2023-806}}
}}}%

\AddToShipoutPictureBG*{%
  \AtPageUpperLeft{%
    \hspace{0.75\paperwidth}%
    \raisebox{-2.5\baselineskip}{%
      \makebox[0pt][l]{\textnormal{FERMILAB-PUB-24-0291-PPD}}
}}}%

\pubyear{\the\year{}}

\begin{document}
\label{firstpage}
\pagerange{\pageref{firstpage}--\pageref{lastpage}}
\maketitle

\begin{abstract}
We report constraints on a variety of non-standard cosmological models using the full 5-year photometrically-classified type Ia supernova sample from the Dark Energy Survey (DES-SN5YR). Both Akaike Information Criterion (AIC) and Suspiciousness calculations find no strong evidence for or against any of the non-standard models we explore. When combined with external probes, the AIC and Suspiciousness agree that 11 of the 15 models are moderately preferred over Flat-$\Lambda$CDM suggesting additional flexibility in our cosmological models may be required beyond the cosmological constant. We also provide a detailed discussion of all cosmological assumptions that appear in the DES supernova cosmology analyses, evaluate their impact, and provide guidance on using the DES Hubble diagram to test non-standard models. An approximate cosmological model, used to perform bias corrections to the data holds the biggest potential for harbouring cosmological assumptions. We show that even if the approximate cosmological model is constructed with a matter density shifted by $\Delta\om\sim0.2$ from the true matter density of a simulated data set the bias that arises is sub-dominant to statistical uncertainties. Nevertheless, we present and validate a methodology to reduce this bias. 
\end{abstract}




\section{Introduction} \label{sec:intro}
Our understanding of the Universe fundamentally changed in the late 1990s with the remarkable discovery that the expansion of our Universe is accelerating \citep{Riess_1998, Perlmutter_1999}. This discovery established $\Lambda$CDM as the standard model of cosmology, which asserts that the Universe at late times is dominated by dark energy in the form of a cosmological constant $\Lambda$ and cold (non-relativistic), pressureless dark matter (CDM). However, the nature of dark energy remains a mystery.

In this paper we use the complete photometrically-classified type Ia supernova (SN Ia) data set from the Dark Energy Survey (DES) -- which represents the largest, most homogeneous SN data set to date -- to assess whether the latest SN Ia data prefers any non-standard cosmological models over $\Lambda$CDM.

While $\Lambda$CDM fits most data well, it lacks a physical motivation and is currently unable to alleviate tensions between early time and late time measurements of parameters such as the current expansion rate of the Universe, $H_0$ \citep{2020_planck, SH0ES_2021}. These two limitations have led to a wealth of exotic cosmological models being proposed \citep[see][for a review]{Di_Valentino_2021}. 

Non-standard cosmological models attempt to explain observations in a variety of ways, ideally with some physical justification. Models that mimic the late time acceleration include dynamical vacuum energy, cosmic fluids, scalar fields as well as modifications to the theory of general relativity. Other models challenge our assumption of large-scale homogeneity and isotropy, and attribute the dimming of distant supernovae to local spatial gradients in the expansion rate and matter density, rather than due to late time acceleration \citep{AlonsoDavid2010Lsss}. 

Previous analyses have shown that many non-standard models are able to explain the current data \citep[e.g.][]{Davis_2007, Sollerman_2009,  Li_2011, 2016_Hu_IDE_SN, Dam_2017, Zhai_2017, lovick2023nongaussianlikelihoodstypeia}, although none have shown strong improvement over $\Lambda$CDM.  In general non-standard models have only been a good fit to the data if they are able to mimic the expansion history of $\Lambda$CDM for some choice of parameters. These analyses conclude that new, more statistically powerful data, across a wide range of cosmological observations are required to discriminate between models.

The Dark Energy Survey was designed to provide such data and to reveal in detail both the expansion history and large-scale structure of the Universe.  Type Ia supernovae are one of the four pillars of DES science, the others being baryon acoustic oscillations \citep[BAO;][]{desbao24}, galaxy clustering \citep{desgc1, desgc2, desgc3}, and gravitational lensing \citep{deswl1, deswl2, deswl3}. 

In this paper we focus on the DES-SN5YR sample \KPp~containing 1829 SNe. The DES-SN5YR sample consists of 1635 SNe from the the full five years of the DES survey, of which 1499 have a machine learning probability of being a type Ia larger than 50 per cent and range in redshift from $0.10$ to $1.13$. This is combined with an external sample of 194 spectroscopically confirmed low-$z$ SNe Ia (see Section~\ref{sec:data}).

Our work builds on previous analyses of non-standard models in two ways. (1) we carefully analyse any cosmological assumptions and approximations that have gone in to the derivation of the information that appears in the Hubble diagram, and estimate their impact. We also provide a prescription for others who would like to use DES SN data to test their own non-standard models, and to provide confidence that there are no hidden assumptions that could bias their result. (2) We constrain a set of non-standard models using the DES-SN5YR sample, with the aim of providing the tightest constraints using SNe Ia measurements alone.

This paper is organised as follows. In Section~\ref{sec:pipeline} we describe the cosmology pipeline used to produce a Hubble diagram, focusing on aspects of the pipeline that contain any cosmological model dependence. In Section~\ref{sec:omw_degen}, we introduce a new parameter, $\mate$, that can be used as a single-number to summarize supernova cosmology constraints in the $w$-$\om$ plane. This new parameter is useful for testing the impact of the reference cosmology used in our simulated bias corrections in Section~\ref{sec:BiasCor} and the fiducial cosmology used while determining the standardised magnitudes of SN Ia in Section~\ref{sec:test_SALT2mu}. Section~\ref{sec:data} describes the data (SN and other external data sets) that we use in this analysis. In Section~\ref{sec:theory} we describe the theory behind the cosmological models we test and present our results. We discuss our results in Section~\ref{sec:discussion} and conclude in Section~\ref{sec:conclusions}.

\section{Cosmology Pipeline}\label{sec:pipeline}

\begin{figure*}
    \centering   \includegraphics[width=\linewidth]{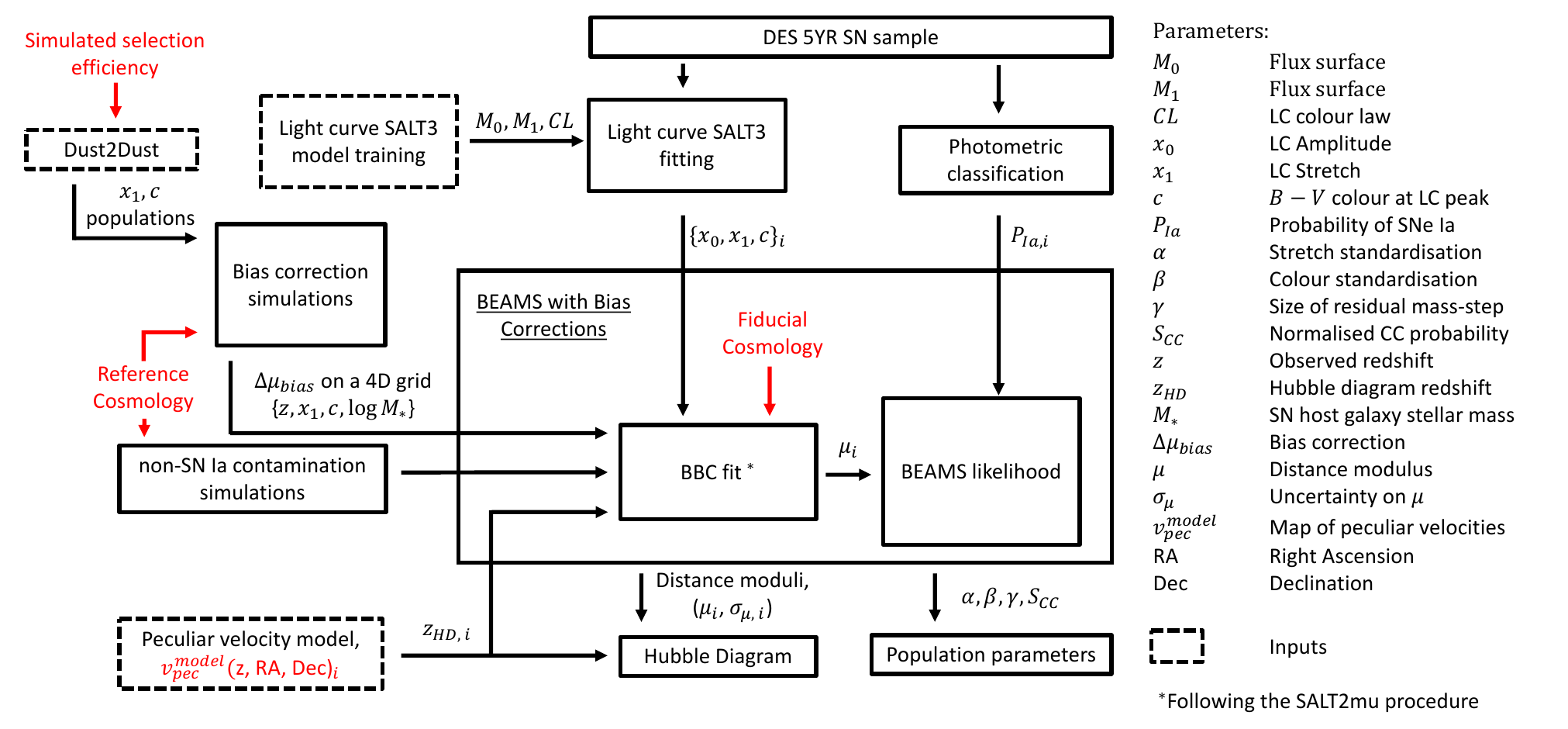}
    \caption{An overview of the light curve to cosmology pipeline.  Here, an emphasis is placed on potential cosmological dependencies (red) and the flow of parameters at each stage. Note that we have also omitted parameter uncertainties and the associated covariances for clarity. However, we have included the final uncertainty term, $\sigma_{\mu, i}$ which includes the intrinsic scatter and a contribution based on the probability of the SN event being a CC contaminant (see Section~\ref{sec:BEAMS}). A subscript $i$ has been added to SN-dependent parameters. Dashed boxes represent external {\sc pippin} inputs.}
    \label{fig:pipeline}
\end{figure*}

Here, we focus on some areas of the DES-SN5YR baseline analysis described in \cite{vincenzi24} --- all the way from light curves to cosmology --- that are, or may appear to be, subject to cosmological dependencies (highlighted in red in Fig.~\ref{fig:pipeline}).  We aim to provide clarity for others who want to use the DES-SN5YR sample to fit their own models. 

The pipeline, illustrated in Fig.~\ref{fig:pipeline}, is run within the {\sc pippin} framework \citep{Hinton2020}, built around several key components 
including the {\sc SALT3} light curve fitting algorithm \citep{Kenworthy_2021}, the {\sc superNNova} photometric classifier \citep{2020MNRAS.491.4277M}, {\sc snana} light curve fitting and simulation for bias corrections \citep{2009_SNANA} producing a bias-corrected Hubble diagram with the ``Beams with Bias Corrections" (BBC) formalism \citep{2017_BiasCor}. We now describe each in turn.

\subsection{Light curve fitting}
 To convert sparse light curve observations to SN standardization parameters we use the {\sc SALT2} model framework \citep{guy_2007, guy_2010} as implemented by the {\sc SALT3} software \citep{Kenworthy_2021}. {\sc SALT3} fits for the time of B-band peak and encapsulates the SN behaviour using three parameters: $x_0$ is the overall amplitude of the light curve; $c$ is related to the $B-V$ colour of the SN at peak brightness; and $x_1$ describes the width of the light curve (stretch). For further details on the light curve fitting used on the DES-SN5YR sample see \cite{2023MNRAS.tmp..345T}. 

 The {\sc SALT3} framework is cosmology independent, except for the assumption that light curves are time-dilated \citep{white24} by a factor of $(1+z_{\rm obs})$. Note that the observed redshift is used to calculate time dilation, therefore there is no peculiar velocity correction at this stage.

\subsection{SN Ia distances}
The distance moduli, $ \mu_{{\rm obs},i}$ of SNe Ia are calculated using the modified Tripp equation \citep{1998_tripp},
\begin{equation} \label{eq:TrippMOD}
\mu_{{\rm obs},i} = m_{x,i} +\alpha x_{1, i}-\beta c_i - \gamma G_{\mathrm{host},i} - \mathcal{M} - \Delta \mu_{\text{bias},i}
\end{equation}
where $m_{x} = -2.5\text{log}(x_0)$; $\mathcal{M}$, is a combination of the SN Ia absolute magnitude, $M$, and the Hubble constant $H_0$, which is marginalised over (see Section~\ref{constraining}); and $\alpha$ \& $\beta$ are nuisance parameters that represent the slopes of the stretch-luminosity and colour–luminosity relations respectively. $\gamma$ is an additional nuisance parameter that accounts for a correlation between standardised SN luminosities and host-galaxy stellar mass, $M_{*}$. This dependency is modelled as a mass-step correction \citep{2010_SNLS3, 2019_DESanalysis}. The final term in equation~(\ref{eq:TrippMOD}), $\Delta \mu_{\text{bias}}$ is applied to each SN to correct for selection effects. 

\subsection{BEAMS with Bias Corrections}\label{sec:BEAMSwBC}
The BEAMS with Bias Corrections \citep[BBC;][]{2017_BiasCor} framework returns a Hubble diagram from a photometrically\footnote{The BEAMS formalism can equivalently be applied to a spectroscopic SNe Ia sample, by setting the probability of each SN event being type Ia to $1$.} identified sample of SNe Ia. It does this by maximising the BEAMS likelihood (Section~\ref{sec:BEAMS}) that accounts for the probability of the SN event being a core-collapse (CC) contaminant while also incorporating bias corrections (Section~\ref{sec:BBC}) and determining global nuisance parameters, $\alpha$, $\beta$ and $\gamma$ from equation~(\ref{eq:TrippMOD}) (Section~\ref{sec:salt2mu}). Therefore, along with the Hubble diagram, BBC also outputs the fitted global nuisance parameters, the uncertainty on the estimated distance moduli, $\sigma_{\mu, i}$, and a classifier scaling factor that is introduced in Section~\ref{sec:BEAMS}.

\subsubsection{The BEAMS likelihood}\label{sec:BEAMS}
Photometric SN samples rely on a classifier to provide a probability of each SN being type Ia or else a contaminant such as core collapse SN or peculiar SN Ia. The DES-SN5YR baseline analysis uses machine learning techniques to classify SN via the open-source algorithm {\sc superNNova} \citep{2020MNRAS.491.4277M}.\footnote{\url{https://github.com/supernnova/SuperNNova}} This classification has no cosmological dependence beyond the assumption that the light curves are time dilated by $(1+\zobs)$.  The predictions of these classifiers, $P_{{\rm Ia}, i}$ are incorporated into the cosmology pipeline by using the \lq Bayesian Estimation Applied to Multiple Species\rq\ (BEAMS) approach \citep{kunz07,Hlozek_2012, kunz2013}. The BEAMS approach, involves maximising the BEAMS likelihood, which includes two terms, one that models the SN Ia population and another that models a population of contaminants. Compared to the traditional likelihood used in spectroscopic samples, the BEAMS likelihood adds one fit parameter, the $P_{\text{CC}}$ scaling factor $S_{\text{CC}}$. The distance uncertainties are then renormalized to ensure that likely contaminants have inflated distance uncertainties and are down-weighted when fitting cosmology. For detailed descriptions of the BEAMS likelihood see \citet{kunz2013}, \citet{2017_BiasCor} and \citet{maria_2021}.  

\subsubsection{Bias corrections}\label{sec:BBC}
The BBC approach uses the BEAMS formalism, and estimates the final term in equation~(\ref{eq:TrippMOD}), $\Delta \mu_{\text{bias}}$, using simulations that model the survey detection efficiency, Malmquist bias as well as other biases introduced in the analysis. Simulations of the DES-SN5YR sample are generated using {\sc snana}\footnote{\url{https://github.com/RickKessler/SNANA}} \citep{2019_DESbias} where light curves are modelled using the {\sc SALT3} framework and the `Dust2Dust' fitting code \citep{P21_dust2dust} measures the underlying population
of stretch and colour, including their correlations with 
host properties.

The simulations used for bias corrections within the baseline analysis are performed using a {\em reference} cosmology of Flat-$\Lambda$CDM with parameters $(\Omega_{m},w)_{\text{ref}} = (0.315, -1.0$). 
There is an underlying assumption in the BBC framework that the bias correction simulations accurately describe the intrinsic properties of the SNe Ia and selection effects. 

The bias correction step thus holds the biggest potential for harbouring cosmological assumptions that could influence the cosmological results.  However, the dependence on the reference cosmology has been shown to be weak for models that have similar\footnote{\cite{2019_DESanalysis} shift the reference cosmology from the best fit by $\Delta w = -0.05$ and find the difference in distance biases are less than 2 mmag across the entire redshift range.} evolution of magnitude versus redshift \citep{2017_BiasCor, 2019_DESanalysis}. Nevertheless, in the analysis of non-standard cosmologies that have the flexibility to deviate significantly from the standard cosmological models, this may no longer be true.  In Section~\ref{sec:BiasCor}, we extend on previous work and quantify the cosmological bias resulting from more extreme reference cosmologies in the context of the DES-SN5YR baseline analysis, and provide a prescription for how to fit models that deviate from the reference cosmology significantly in their evolution of magnitude versus redshift. 

\subsubsection{BBC fit}\label{sec:salt2mu}
The global nuisance parameters, $\alpha$, $\beta$ and $\gamma$ are used to standardise SN magnitudes and are determined using the BBC fitting algorithm (which has previously been referred to as {\sc SALT2mu}), following the redshift binning procedure in \cite{Marriner_2011} and equation 3 of \citet{2017_BiasCor}. BBC employs a {\em fiducial} cosmology\footnote{Note that the fiducial cosmology used within the BBC fit in general can differ from the reference cosmology used to simulate SNe used for bias corrections.}  that provides an arbitrary smooth Hubble diagram in each redshift bin. BBC fits for $\alpha$, $\beta$ and $\gamma$ by minimizing the Hubble residuals to the fiducial cosmology among $N_b=20$ logarithmically-spaced redshift bins as well as fitting for a magnitude offset in each bin. 

\begin{figure}
    \centering
    \includegraphics[width=0.8\linewidth]{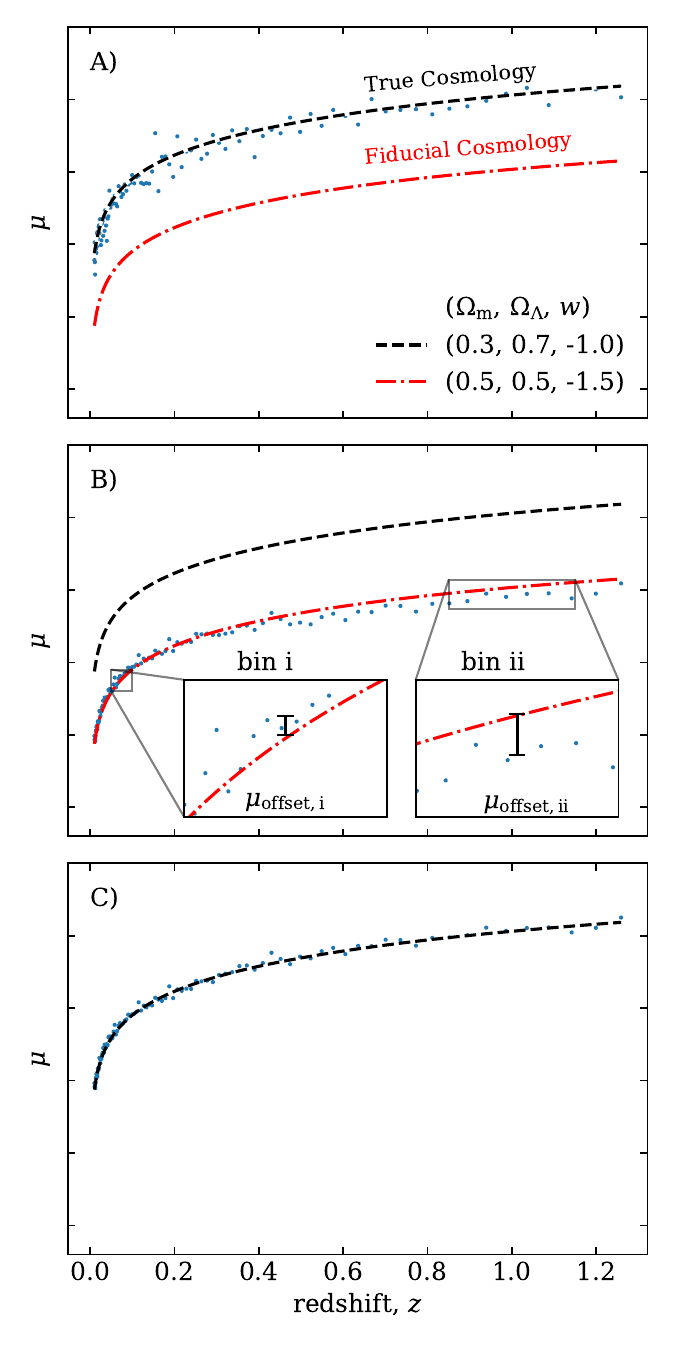}
    \caption{A visualization of the BBC fit. A) We start with SN distance moduli that are not standardized and therefore have large scatter, here the true cosmology is shown as a black dashed line. BBC employs a fiducial cosmology (red dot-dashed line) that in general is different from the true cosmology. In B) we then fit for $\alpha$ and $\beta$ by minimizing the residual to the fiducial cosmology while simultaneously fitting for magnitude offsets in $N_b=20$ logarithmically spaced redshift bins. The insets show the varying size of the offsets in different bins relative to the average offset, $\mu_{\mathrm{offset, }b} = \Delta \mu_b - \mu_{\mathrm{avg}}$. As $\mu_{\mathrm{offset, i}}$ does not in general equal $\mu_{\mathrm{offset, ii}}$ this procedure allows the data to better resemble the true cosmology (black dashed line) approximately mapping the fiducial cosmology on to the true one by quantifying how much the observations deviate from the fiducial cosmology in each redshift bin. In B), the data has been shifted to the fiducial cosmology for illustrative purposes and in C) we shift the data back. Therefore, for this example, $\mu_{\mathrm{avg}}$ would be positive (the data actually sits above our fiducial cosmology), however $\mu_{\mathrm{offset, }i}$ and $\mu_{\mathrm{offset, }ii}$ would be positive and negative respectively (because the data sits above and below the average offset respectively). While this example is exaggerated it is useful to provide insight into BBC and highlight that the method has minimal cosmological dependence.}
    \label{fig:SALT2muVisual}
\end{figure}

The default fiducial cosmology used in the BBC fit, for the DES-SN5YR analysis, is the Flat-$\Lambda$CDM model with parameters $(H_0, \Omega_{\text{m}}) = (70, 0.3)$. This choice may cause confusion within the community regarding a potential cosmology dependence. Fig.~\ref{fig:SALT2muVisual} provides an exaggerated visualization of the BBC fit to show i) fitting for magnitude offsets in redshift bins allows the data to better resemble its naturally standardized state (with $\alpha_{\mathrm{fit}}, \beta_{\mathrm{fit}}$ consistent with the true values); ii) the magnitude offsets (approximately) map the fiducial cosmology on to the true one by quantifying how much the observations deviate from the reference cosmology in each redshift bin; and iii) that this procedure removes the dependence on cosmological parameters.

\citet{Marriner_2011} show that the fit for $\alpha$ and $\beta$ is decoupled from the choice of fiducial cosmology if the number of redshift bins is sufficiently large. Furthermore, \citet{KVA23} performs a limited study that looks at the standard deviation of the Hubble residuals of the BBC fit \citep[see Table 1 of][]{KVA23}. In Section~\ref{sec:test_SALT2mu}, we re-test this result and extend on the work of \citet{Marriner_2011} and \citet{KVA23} by explicitly testing extreme cosmologies as well as showing that the impact on cosmology-fitted parameters is negligible. Finally, we present an alternate approach that does not require a fiducial cosmology and achieves consistent fits for $\alpha$ and $\beta$.

\subsubsection{SN Ia distance uncertainties}\label{sec:uncertainties}
Following the Pantheon+ analysis \citep{2022_pantheon_analysis}, the distance modulus uncertainties $\sigma_{\mu,i}$ are calculated within the BBC approach as,
\begin{equation}
\begin{split}
        \sigma_{\mu,i}^2 = f(z_i, c_i, M_{*,i})  &\sigma_{\mathrm{S3fit},i}^2 + \sigma^{2}_{\mathrm{floor}}(z_i, c_i, M_{*,i}) \\
   &+ \sigma_{\mathrm{lens},i}^2 + \sigma_{z,i}^2 + \sigma_{{\rm vpec},i}^2
   \end{split}
\label{eq:sig1}
\end{equation}
 where $\sigma_{\mathrm{S3fit},i}$ includes the uncertainties on the light curve parameters and the associated covariances; while $\sigma_{\mathrm{lens},i}$  and $\sigma_{z,i}$ are uncertainties associated with lensing effects and spectroscopic redshifts, respectively.  $f(z_i, c_i, M_{*,i})$ and $\sigma_{\mathrm{floor}}(z_i, c_i, M_{*,i})$ are survey-specific scaling and additive factors that are estimated from the BBC simulations. Finally, $\sigma_{{\rm vpec},i}$ accounts for uncertainties due to peculiar velocities, including both uncertainties in linear-theory modelling and non-linear unmodelled peculiar velocities, as discussed in Sec~\ref{sec:vpecmodel}.

\subsection{Modelling peculiar velocities}\label{sec:vpecmodel}
The redshift that is compared to SN distances should be entirely due to the expansion of the universe. However, in practice the redshift that we measure contains contributions due to peculiar velocities of the SN and its host galaxy. The DES-SN5YR baseline analysis uses peculiar velocities presented by \cite{Peterson_2022}, which are determined from the 2M++ density fields \citep{2015MNRAS.450..317C} with global parameters and group velocities used from \citet{2020MNRAS.497.1275S} and \citet{2015AJ....149..171T} respectively and a 240 km s$^{-1}$ uncertainty on these estimates. While the determination of the peculiar velocity corrections includes a fiducial cosmology, the corrections have the largest impact at low redshifts where the cosmology dependence is negligible. Although \cite{Peterson_2022} show that the impact of peculiar velocity corrections on $H_0$ and $w$ fits are at the 1 per cent level, the impact of the fiducial cosmology in the derivation of those corrections is negligible compared to the uncertainty in the peculiar velocity map, and therefore we do not consider it further in this work.

\section{The \texorpdfstring{$\Omega_{\lowercase{\mathrm{m}}}-\lowercase{w}$}{om-w} degeneracy}\label{sec:omw_degen}

There is a degeneracy between the equation of state of dark energy and the matter content of the universe for distance indicators within generalised dark energy models. It has long been known that this degeneracy makes it more difficult to assess systematics on $\Omega_{\rm m}$ and $w$ separately.

Large shifts in the best fit parameters may not be significant if they occur along the degeneracy direction, but the same size shifts could be very significant if they occur perpendicular to the degeneracy direction.  In the DES cosmology analysis we use two methods to account for that degeneracy.  The first is setting a prior on matter density\footnote{Either a CMB-like prior or a direct matter density prior.} and only considering changes in $w$, the other is testing a new parameter $\mate(z)$ that allows us to present a single non-degenerate number summarising a SN Ia constraint in the $w$-$\om$ plane.

To link Flat-$w$CDM and cosmography, we can use the acceleration equation
\begin{equation}\label{eq:acceleration}
    \frac{\Ddot{a}}{a} = -H_0^2 \frac{1}{2}\left[ \Omega_{\mathrm{m}} a^{-3} + \Omega_{\rm de} \left(1+3w\right) a^{-3(1+w)} \right],
\end{equation}
where $\Omega_{\rm de} = 1 - \Omega_{\mathrm{m}}$ for a spatially flat universe. Note that $H\equiv\dot{a}/a$, therefore using the definition of the deceleration parameter, $q\equiv-\ddot{a}/(a H^2)$ we can rearrange equation~(\ref{eq:acceleration}) and express $q(H/H_0)^2$ as a function of the energy mix of a Flat-$w$CDM universe,
\begin{equation}
    \mate(z) = \frac{1}{2}\left[ \Omega_{\mathrm{m}} a^{-3} + \Omega_{\rm de} \left(1+3w\right) a^{-3(1+w)} \right]
\end{equation}
where we have defined $\mate \equiv -\ddot{a}/(aH_0^2)\equiv q(H/H_0)^2$ and $a = (1+z)^{-1}$.

\begin{figure}
    \centering
    \includegraphics[width=\linewidth]{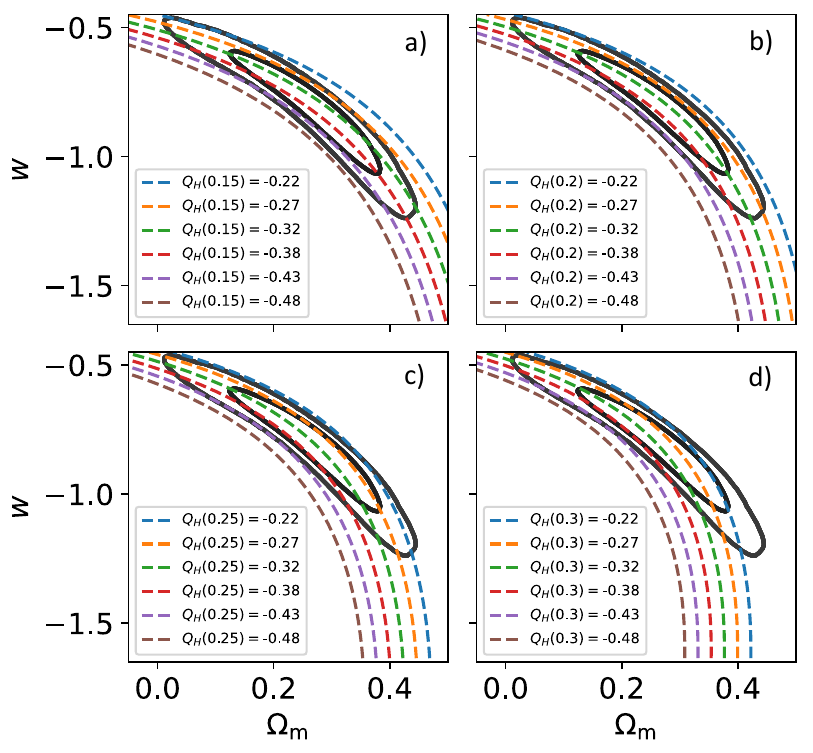}
    \caption{Comparing lines of constant $\mate(z)$ with $z = 0.15, 0.20, 0.25, 0.30$ for panels a), b), c), d) respectively. Here, we overlay in each panel the Flat-$w$CDM $1\sigma$ and $2\sigma$ contours for the DES-SN5YR sample.}
    \label{fig:BBC_bias_compare}
\end{figure}
In Fig.~\ref{fig:BBC_bias_compare} we show lines of constant $\mate(z)$ overlaid on to the $1\sigma$ and $2\sigma$ contours for the DES-SN5YR sample. Since the $\mate(z)$ parameter is redshift dependent, it is not as universal as a parameter such as $S_8=\sigma_8\sqrt{\om/0.3}$, which defines a quantity that is relatively independent of the $\sigma_8$ and $\om$ degeneracy in lensing studies.  Instead, we can select a redshift that matches the degeneracy direction of the sample. In the top right subplot of Fig.~\ref{fig:BBC_bias_compare} we show that $\mate(0.2)$ makes a good approximation for the $w$-$\om$ degeneracy line for the DES-SN5YR sample. Using the $\mate(0.2)$ parameter, we can therefore use a single number to approximate the DES-SN5YR constraints on the Flat-$w$CDM model and find $\mate(0.2)=-0.340\pm0.032$ (which includes statistical and systematic uncertainties).

\begin{figure*}
    \centering
    \includegraphics[width=\linewidth]{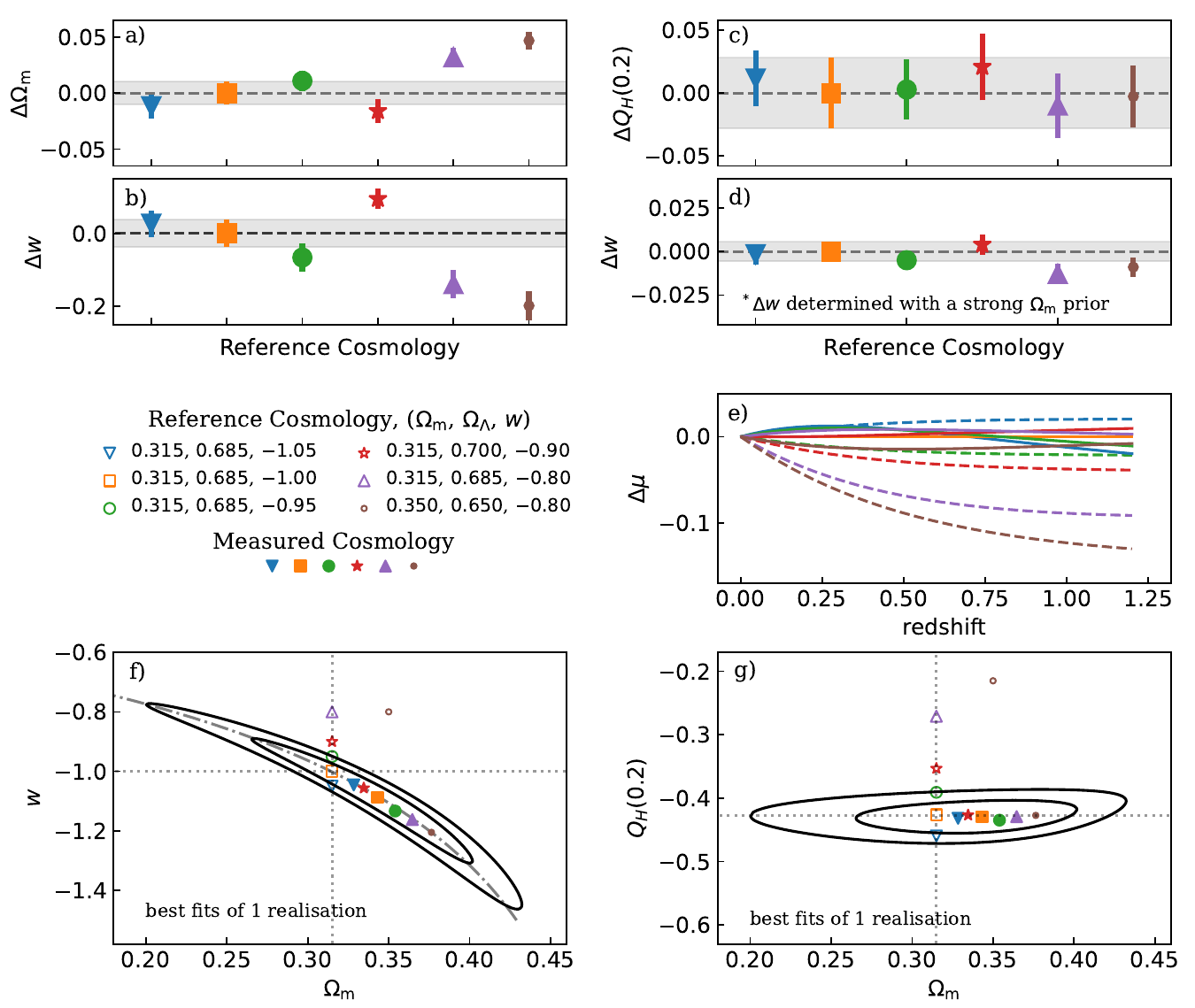}
    \caption{a) and b): Shifts in $\Omega_{\mathrm{m}}$ and $w$ (solid points) when using different BBC simulations that are distinguished by a unique reference cosmology (shown by open symbols; listed in the figure legend). The shifts are measured relative to the perfect scenario (orange square) where the reference cosmology is equal to the true cosmology of our simulated data. c) and d): The associated mean shifts in $\mate(0.20)$~(with no prior) as well as $w$ determined with a strong prior on the matter density of $\Omega_{\mathrm{m}}=\Omega_{\mathrm{m, true}}\pm 0.001$, which minimizes the impact that the $\Omega_{\mathrm{m}}-w$ degeneracy has on investigating the BBC reference cosmology. For panels a) - d) we have averaged over 25 DES-SN5YR simulations. Note also that the error bars show the uncertainty on the shift in the mean -- {\em not} the uncertainty on the parameters, which is larger. e): Calculated residual distance moduli of the reference cosmologies (dashed lines) relative to the baseline cosmology $(\Omega_{\mathrm{m}}, w) = (0.315,-1.0)$ in orange. The solid lines represent the variation in the expansion history from the perfect scenario using the mean of the best fit parameters. f): Best fit parameters (solid points) for \textit{1 realisation} of simulated data determined using a unique BBC reference cosmology (shown by open symbols). The  $1\sigma$ and $2\sigma$ contours shown are for the ideal case (orange square). The grey dotted dashed line represents the $\mate(0.2)$ parameter. g): Equivalent information to that contained in plot f) but converted to $\Omega_{\mathrm{m}}-\mate(0.2)$ space.
    }
    \label{fig:Combinedshifts}
\end{figure*}

Changes to the analysis that only cause shifts along the degeneracy direction have a very small effect on $Q_H$ even though they can have a misleadingly large effect on $\om$ and $w$ (misleading since those shifts are strongly correlated). $Q_H$ is thus an excellent measure by which to evaluate the impact of analysis choices on the supernova cosmology results (see Fig.~\ref{fig:Combinedshifts}).

\section{Reference cosmology in the bias correction simulations} \label{sec:BiasCor}

\cite{2017_BiasCor} show that any dependence on the reference cosmology is weak when the reference cosmology is similar to the true evolution of magnitude versus redshift \citep[see Sec~6.1 and Fig.~7. of][for details]{2017_BiasCor}. Here, we reevaluate this systematic and also show that using a reference cosmology even $10\sigma$ away from the true cosmology has less than a $1\sigma$ shift in the results. We also present an iterative method that can be used to reduce even that small systematic offset. 

\subsection{Testing the impact of the reference cosmology}\label{sect:testingref}
To examine the impact that the reference cosmology used for the bias correction simulations has on our cosmology fits, we generate and analyze 25 realisations of simulated data. These are created with a Flat-$w$CDM cosmology with parameters $(H_0, \Omega_{\mathrm{m}}, w) = (70,0.315,-1.0)$.  We also generate six different BBC simulations, each with a unique reference cosmology. For comparison, in Fig.~\ref{fig:Combinedshifts}e we plot each reference cosmology (dashed lines) relative to the cosmology used to generate our simulated data (orange). 

The average shifts in $\Omega_{\mathrm{m}}$ and $w$ from the perfect scenario in which the reference cosmology is equal to the true cosmology of our simulated data are shown in Fig.~\ref{fig:Combinedshifts}a and Fig.~\ref{fig:Combinedshifts}b respectively.\footnote{We note that these biases appear larger than those found by \cite{2017_BiasCor} because they used a strong $\Omega_{\mathrm{m}}$ prior, which is more similar to what we show on the lower panel of the top-right plot. We discuss why these larger shifts are not concerning below.} 

In Fig.~\ref{fig:Combinedshifts}f we plot the results in the $w-\om$ plane for a single realisation.  The contours and solid orange square are for the ideal case in which the reference cosmology matches the true cosmology.  The other symbols show the results when using different reference cosmologies, where the open symbols show the input reference cosmology and the solid symbols show the resulting best fit parameters.    

This shows that while the shifts in $\om$ and $w$ seem large, when viewed in 2D parameter space they all fall along the $\Omega_{\mathrm{m}}-w$ degeneracy direction and are thus all well within $1\sigma$.

The dot-dashed line in Fig.~\ref{fig:Combinedshifts}f shows the $\mate(0.2)$ parameter, representing the degeneracy line.  Note that the ideal redshift for $Q_H$ to match the degeneracy direction will change depending on the data set. In Fig.~\ref{fig:Combinedshifts}c we plot the average shift in $\mate(0.2)$ and in Fig.~\ref{fig:Combinedshifts}d we plot the shift in $w$ after applying a strong prior on the matter density $\Omega_{\mathrm{m}}=\Omega_{\mathrm{m, true}}\pm 0.001$. The fact that the shifts in $\mate(0.2)$ and $w|_{\Omega_{\mathrm{m, true}}\pm 0.001}$ are negligible shows that the impact of the reference cosmology is small and limited to the degeneracy direction, in agreement with the results from  \cite{2017_BiasCor}.

\begin{figure}
    \centering
    \includegraphics[width=0.9\linewidth]{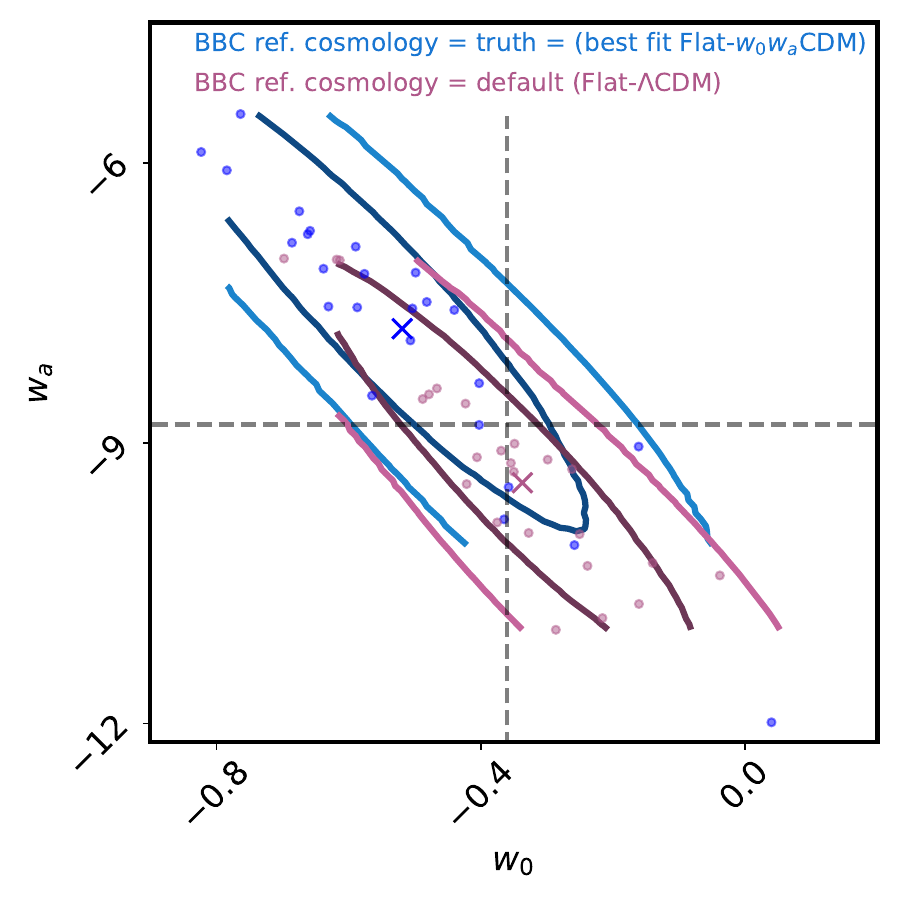}
    \caption{Comparison of the best fit $w_0-w_a$ points (with a prior on the matter density, $\Omega_{\mathrm{m}}=\Omega_{\mathrm{m, true}}\pm 0.001$) determined using the DES-SN5YR baseline reference cosmology (purple) and when the reference cosmology is set to the input cosmology of the simulations (blue). The points show the maximum likelihood values for each realisation and the crosses represent the averages of the those maximum likelihood values. The ellipses are the 1- and 2$\sigma$ contours representing the dispersion of best fit points.}
    \label{fig:w0wa_test}
\end{figure}

We also performed two additional tests that are the inverse of those performed above. Instead of varying the reference cosmology, we fixed the reference cosmology to the baseline cosmology used in the DES-SN5YR analysis and generated 25 realisations of simulated data using both (a) Flat-$w$CDM cosmology with parameters $(H_0, \Omega_{\mathrm{m}}, w) = (70,0.350,-0.8)$ and (b) Flat-$w_0 w_a$CDM cosmology with parameters $(H_0, \Omega_{\mathrm{m}}, w_0, w_a) = (70,0.495,-0.36, -8.8)$. These cosmologies were chosen to match the $\sim 10\sigma$ offset brown point in Fig.~\ref{fig:Combinedshifts} and the best fit Flat-$w_0 w_a$CDM result in the DES-SN5YR analysis respectively. The results are given in Table~\ref{tab:add_tests}. 
For test (a), we again find that the impact of using the incorrect reference cosmology is negligible. For test (b), we see larger shifts in cosmological parameters. However, in this case, their is an additional degeneracy between $w_0-w_a$ that is not accounted for when applying the prior on $\Omega_{\mathrm{m}}$. To visualise this, we plot the 25 realisations in Fig.~\ref{fig:w0wa_test}, which, shows that the best fit points are aligned along the degeneracy line and consistent with the truth. We also note that the uncertainties given in Table~\ref{tab:add_tests} are on the shift in the mean. The shifts are $\Delta w_0 = 0.18\pm0.28$, $\Delta w_a = -1.6\pm2.2$ when using the uncertainty on the parameters.
\begin{table}
\caption{Shifts in the best fit parameters using the DES-SN5YR baseline reference cosmology, from the perfect scenario where the reference cosmology is equal to the cosmology used to generate the simulated data. Here, the uncertainties are on the shift in the mean -- {\em not} the uncertainty on the parameters, which is larger.}
\label{tab:add_tests}
\renewcommand{\arraystretch}{1.3}
\begin{tabular}{lccc}
\hline \hline
Model$^{*}$ & & &  \\
($\Omega_{\mathrm{m}}$, $w_0$, $w_a$)& 
$\Delta Q_\mathrm{H}(0.20)$ &
$\Delta w_0^{\dag}$ &
$\Delta w_a^{\dag}$ \\\hline 
$(0.350, -0.80, 0)$ &  $0.02 \pm 0.05$ &  $0.000 \pm 0.008$ & -  \\ 
$(0.495, -0.36, -8.8)$  & - & $0.18\pm0.06$ &  $-1.6\pm0.4$\\ 
\hline
\end{tabular}
{\par \raggedright $^{*}$  Model used to generate the 25 realisations of simulated data \par}
{\par \raggedright $^{\dag}$  Determined used a prior on  the matter density of $\Omega_{\mathrm{m}}=\Omega_{\mathrm{m, true}}\pm 0.001$.\par}
\end{table}

In summary, this result validates that the BBC baseline approach used in \KP~is able to return a Hubble diagram that represents the true distance versus redshift relation to within $1\sigma$ even given a reference cosmology that is $\sim 10\sigma$ from the truth (brown point in Fig.~\ref{fig:Combinedshifts}) or varies by $\sim \Delta \mu = 0.15$ (brown dashed line in Fig.~\ref{fig:Combinedshifts}e). The apparent bias observed in Fig.~\ref{fig:Combinedshifts}a and Fig.~\ref{fig:Combinedshifts}b is due to showing shifts in degenerate parameters separately, without considering the combined influence on the distance versus redshift relation. Importantly, we can be confident in our bias corrections if the expansion history of a non-standard cosmological model falls within the region bounded by the blue and brown dashed lines in Fig.~\ref{fig:Combinedshifts}e.

\subsection{The iterative method}\label{sec:iterative}
Section~\ref{sect:testingref} validates the procedure used in the DES-SN5YR baseline analysis, showing that the reference cosmology has a small impact on the cosmological results relative to the statistical uncertainties. However, the BBC reference cosmology may become a dominating systematic for future surveys such as the Rubin Observatory's LSST, which will include hundreds of thousands of well measured SNe Ia \citep{LSST_2009}. Furthermore, Fig.~\ref{fig:Combinedshifts} shows that in the case where one finds a tension with other data sets {\em at the extreme ends of the degeneracy direction} (e.g. if the CMB contours were at the top left or bottom right in Fig.~\ref{fig:Combinedshifts}f), it {\em would} be beneficial to ensure a close match to the reference cosmology.  Since we performed a blind analysis, we did not know whether there would be a discrepancy between the BBC reference cosmology and the final fitted cosmology results. We therefore prepared the following method to correct the reference cosmology if
the discrepancy was significant.

\begin{figure}
    \centering
    \includegraphics[width=0.9\linewidth]{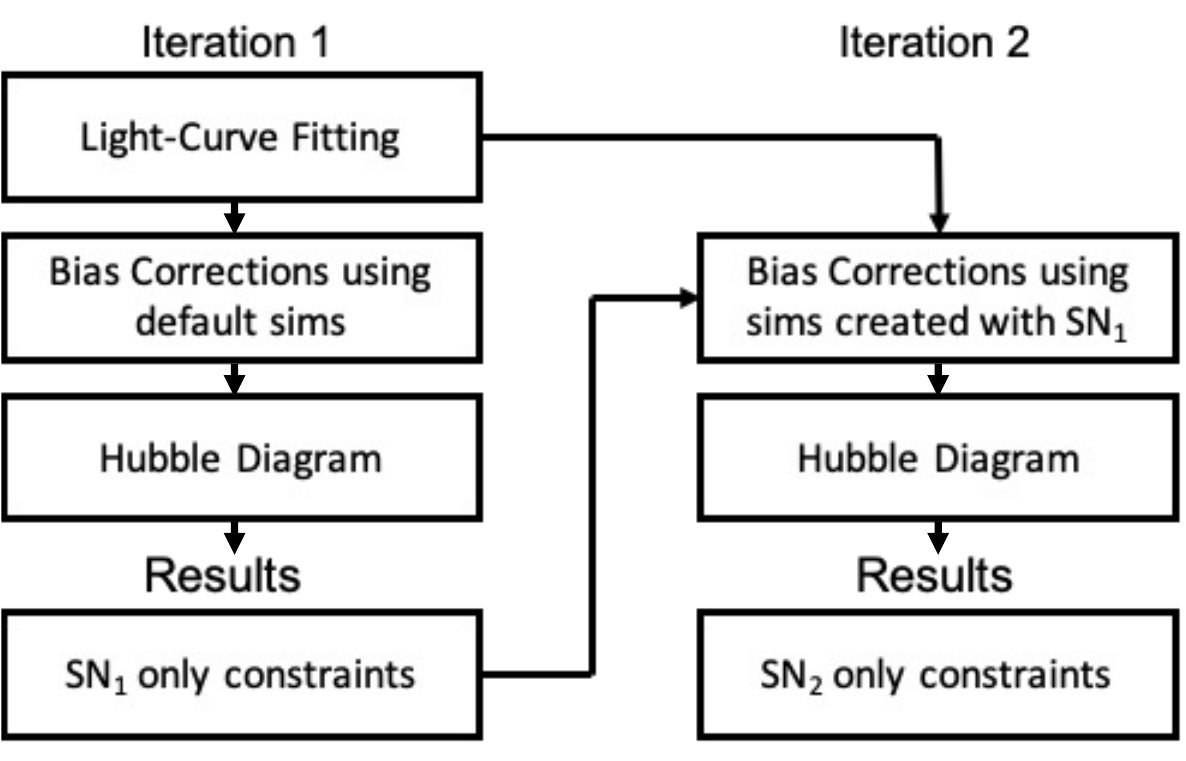}
    \caption{Iterative procedure methodology. During the first iteration, bias corrections are modelled using simulations created using the default reference cosmology with a fixed set of Flat-$w$CDM parameters $\Omega_{\mathrm{m},\text{ref}} = 0.3$ and $w_{\text{ref}} = -1.0$. In the second iteration, the simulations are instead created using the maximum likelihood estimates from the first iteration.}
    \label{fig:analysis_method}
\end{figure}

It was suggested by \cite{2017_BiasCor} that an iterative procedure can be applied where $w_{\text{ref}}$ is updated with the previous $w_{\text{fit}}$ value, to reduce this bias. This procedure is summarised in Fig.~\ref{fig:analysis_method}.  In this work, we test the iterative method by applying it to 10 realisations of simulated data created with a Flat-$w$CDM cosmology with parameters $(H_0, \Omega_{\mathrm{m}}, w)=(70, 0.350, -0.8)$. This cosmology was selected due to its location in parameter space, which is approximately perpendicular to the $\Omega_{\mathrm{m}}-w$ degeneracy line in the direction of a general CMB prior and lies outside a $2\sigma$ region (based on DES-SN5YR simulations) of the default BBC reference cosmology.\footnote{Flat-$\Lambda$CDM with $\Omega_{\mathrm{m}}=0.315$.} 
Table~\ref{tab:iteration_tests} shows the weighted average shift in cosmological parameters from the truth after 10 realisations. Note that the $\Omega_{\mathrm{m}}$ prior was only applied on our final results and was not used during the iterative process. We report both $\Delta w|_{\Omega_{\mathrm{m, true}}\pm 0.001}$ and $\Delta Q_H(0.2)$ and find that both are closer to the truth after applying the iterative method. In particular, we find that $w|_{\Omega_{\mathrm{m, true}}\pm 0.001}$ has shifted by $0.006$ and $Q_H(0.2)$ has shifted by $0.008$ closer to the truth.

We note a limitation of this work that we have not explicitly shown the iterative method converges (because repeatedly redoing the simulations is computationally intensive). However, we performed a third iteration on two random realisations and found that the iterative method remained stable.

 The iterative method was not implemented in the current DES results, because after unblinding we found the best fit cosmology to be sufficiently close to the reference cosmology so as to make any bias insignificant (in Sec.~\ref{sect:testingref} we found $\Delta w \sim 0.01$ given a reference cosmology 10$\sigma$ from the truth). Nevertheless, we conclude that iterating the reference cosmology is a viable method to reduce this bias for future analyses where the reference cosmology may become a dominating systematic. 
\begin{table}
\caption{Testing the iterative method (Section~\ref{sec:iterative}): Weighted average (over 10 realisations$^{*}$) difference in $w$ and $Q_H$ from the truth for the first and second iterations.}
\label{tab:iteration_tests}
\renewcommand{\arraystretch}{1.3}
\begin{tabular}{lcccc}
\hline \hline
Method  &
$\Delta w^{\dag}_{\mathrm{fit-true}}$  &
$\sigma_{w\mathrm{, avg}}$  & 
$\Delta Q_\mathrm{H, fit-true}$ & 
$\sigma_{Q_\mathrm{H, avg}}$ \\ \hline 
\hspace{0mm} Nominal &  -0.023 &  0.028 &-0.051& 0.019  \\ 
\hspace{0mm} $2^{\mathrm{nd}}$ Iteration  &-0.017 &0.025&-0.043&0.019\\ 
\hline
\end{tabular}
{\raggedright $^{*} \Omega_{\mathrm{m}}=0.350$ and $w=-0.8$ was used as the true cosmology. \par}
{\raggedright $^{\dag}$ With a prior on the matter density of $\Omega_{\mathrm{m}}=0.350\pm 0.001$ \par}
\end{table}

\section{Tests of cosmology dependence within the BBC fit}\label{sec:test_SALT2mu}

\begin{figure*}
    \centering
    \includegraphics[width=0.9\linewidth]{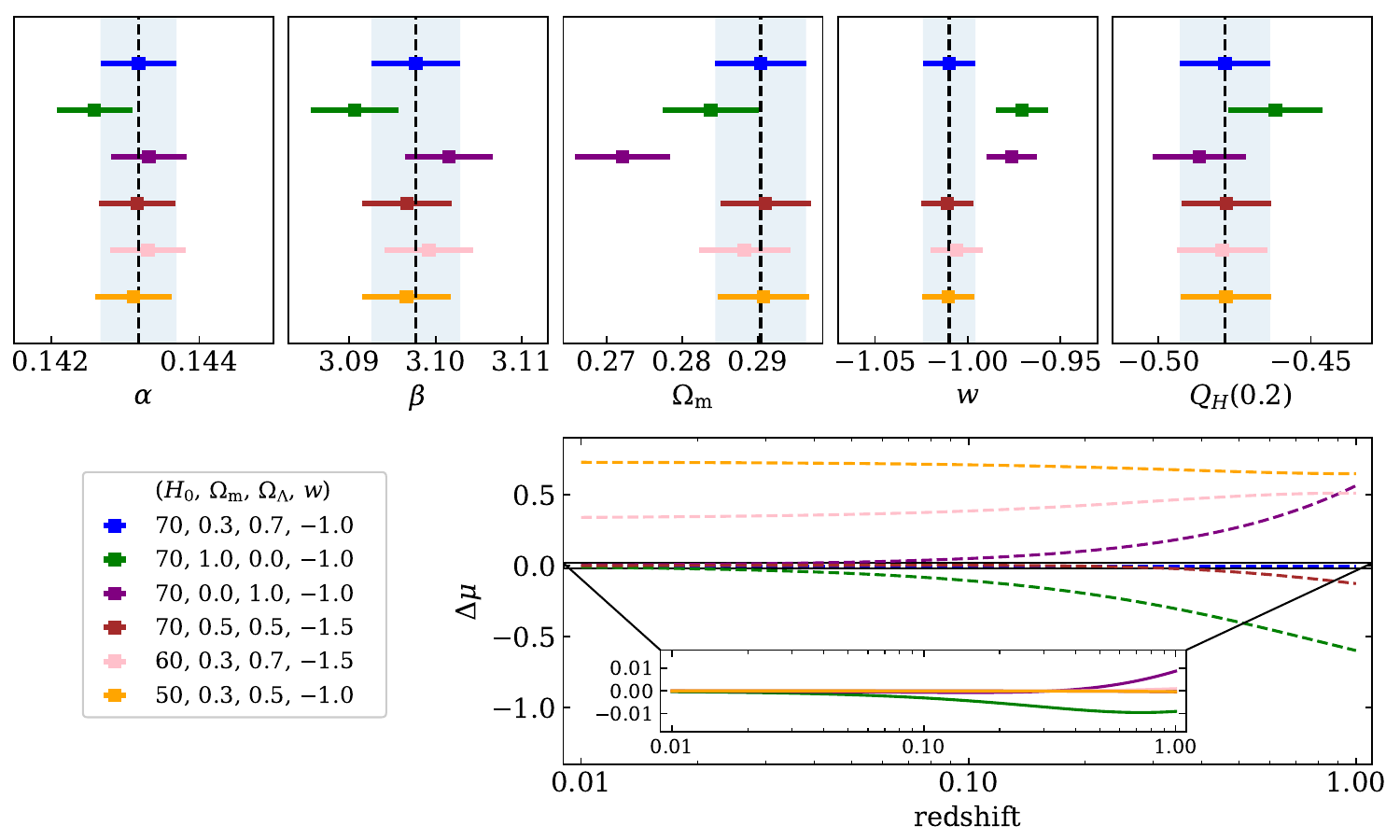}
    \caption{Top panels: Shifts in the average maximum likelihood $\alpha$, $\beta$, 
    $\Omega_{\mathrm{m}}$, $w$ and $Q_H(0.2)$ values after varying the fiducial cosmology within the BBC fit (Section~\ref{sec:test_SALT2mu}). The error bars used are the standard error of the mean and are therefore much larger for the individual case. The values are shown relative to the ideal case (black dashed line) where the fiducial cosmology is equal to the true cosmology used to simulate the data. Only the model with zero matter density, and pure cosmological constant (plum) shows a more than $1\sigma$ shift from the fidicual, and comparison with both the $Q_H$ panel and Fig.~\ref{fig:100_combined} shows that shift is along the degeneracy direction. Bottom right: Variation in the evolution of magnitude versus redshift from the ideal case for (a) the input fiducial cosmological parameters (given in the legend) shown as dashed lines and (b) using the mean of the best fit parameter values shown in the zoomed inset axes as solid lines. }
    \label{fig:Fid_test_cosmoabmean}
\end{figure*}

\begin{figure*}
\centering
\includegraphics[width=\linewidth]{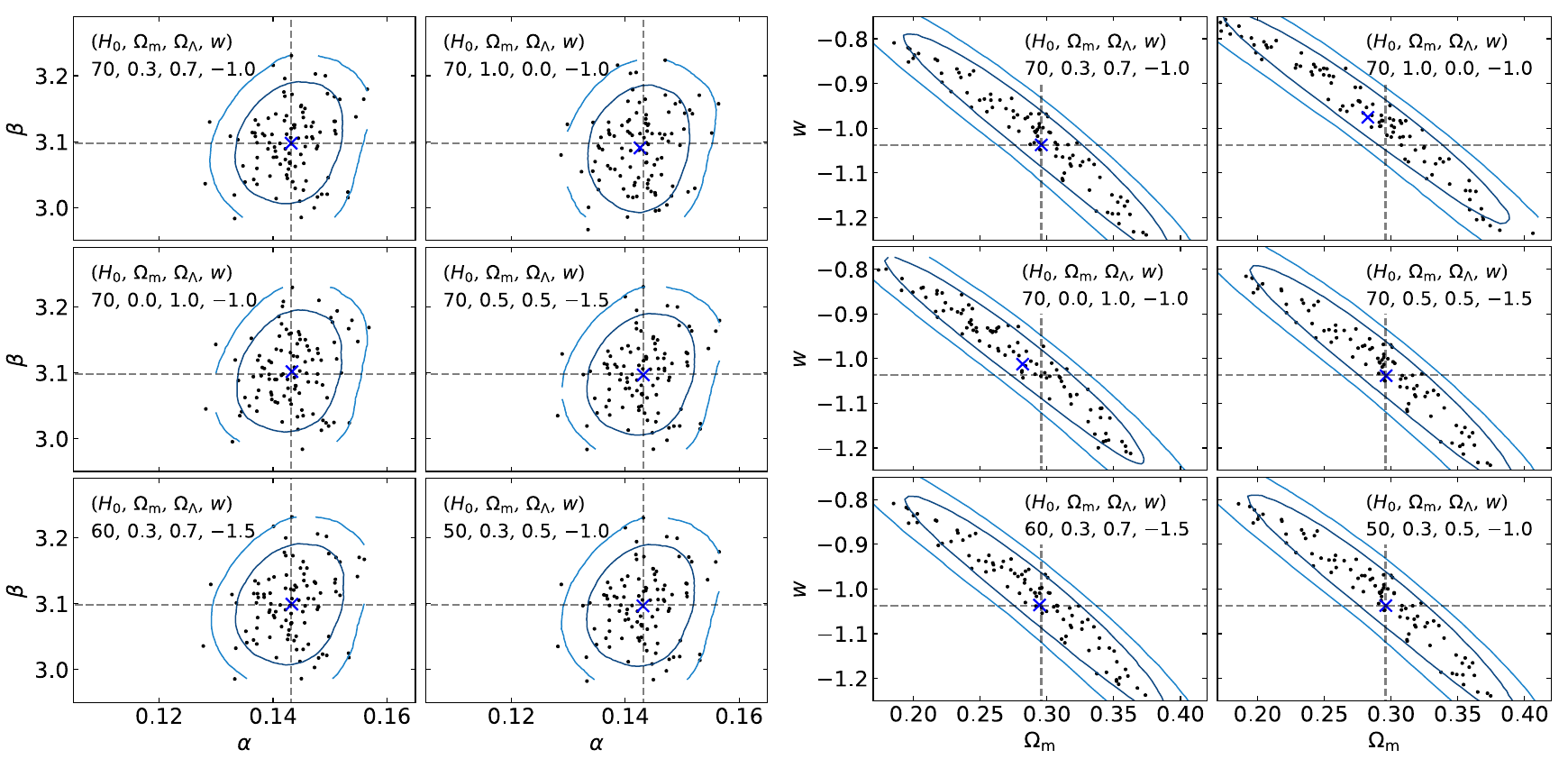}
\caption{Left: The best fit $\alpha$ and $\beta$ for 100 mock realisations for each of six different reference cosmologies as per the legend (see Section~\ref{sec:test_SALT2mu}). The black points show the maximum likelihood values for each realisation and the blue crosses represents the averages of the those maximum likelihood values. The blue ellipses are the 1- and 2$\sigma$ contours representing the dispersion of best fit points. The upper left sub figure represents the perfect scenario where the fiducial cosmology is equal to the true cosmology used to simulate the data. The black dashed lines are used to compare each figure to this ideal case. Right: The equivalent figure after fitting for cosmological parameters, $\Omega_{\mathrm{m}}$ and $w$.}
\label{fig:100_combined}
\end{figure*}
In this section we validate the baseline analysis assumption that the fit for nuisance parameters is decoupled from the choice of fiducial cosmology using 20 logarithmically space redshift bins (for these tests we restrict ourselves to $\alpha$ and $\beta$). 

In total, we generated 100 statistically independent realisations that resemble the DES-SN5YR sample in a spatially Flat-$\Lambda$CDM universe with parameters $(H_0, M_B, \Omega_{\mathrm{m}}) = (70, -19.253, 0.3)$. We ran all 100 realisations through the entire \verb|PIPPIN| pipeline six times with each run distinguished uniquely by the choice of fiducial cosmology within the BBC fitting procedure. The choice of fiducial cosmologies was chosen such that they vary significantly in the evolution of magnitude versus redshift and are shown in the bottom panel of Fig.~\ref{fig:Fid_test_cosmoabmean}.

The left panel of Fig.~\ref{fig:100_combined} compares the maximum likelihood $\alpha$ and $\beta$ values for each of the 100 realisations. The top left sub-plot represents the ideal case where the fiducial cosmology is equal to the true cosmology used to simulate the data. Here, we show how the averages of the 100 maximum likelihood values (blue crosses) compared to the true values (black dashed lines). We also make the equivalent comparison after fitting for cosmological parameters, shown in the right panel of Fig.~\ref{fig:100_combined}. In Fig.~\ref{fig:Fid_test_cosmoabmean} we present the shifts in the average of the maximum likelihood $\alpha$, $\beta$, $\Omega_{\mathrm{m}}$, $w$ and $Q_H(0.20)$ values as a result of varying the fiducial cosmologies within the BBC fit. We also show how the shifts in cosmological parameters impacts the evolution of magnitude versus redshift relative to the ideal case.

We find that the determination of the global nuisance parameters, $\alpha$ and $\beta$, has a weak dependence on the choice of fiducial cosmology; these results are in agreement with those by \cite{Marriner_2011}. Extending on the work by \cite{Marriner_2011}, Fig.~\ref{fig:Fid_test_cosmoabmean} shows that the BBC fit is able to recover the ideal cosmological parameters with less than a $1\sigma$ tension of the standard error given 100 realisations even when using extreme fiducial cosmologies. The two fiducial cosmologies that result in the largest shift in cosmological parameters are unsurprisingly also the two cosmologies that deviate the most in the slope of the distance versus redshift relation 
$(H_0, \Omega_{\mathrm{m}}, \Omega_{\Lambda}, w)=(70, 0.0, 1.0, -1.0)$ and $(70, 1.0, 0.0, -1.0)$. However, both the $Q_H(0.2)$ panel and Fig.~\ref{fig:100_combined} show that shift is along the degeneracy direction.

Finally, the lower right of Fig.~\ref{fig:Fid_test_cosmoabmean} shows the $\mu$ differences between the fiducial cosmologies (dashed lines) and even shifts of $\mu$ up to $0.5$ across the $z$-range have negligible impact on the best fit expansion history (solid lines).

\subsection{Is a fiducial cosmology required?}\label{sec:isfidreq}
Often, the role of the fiducial cosmology within the BBC fit causes confusion -- both because of perceived cosmology dependence (which we have shown is negligible for any reasonable cosmology in Section~\ref{sec:test_SALT2mu}) and because it is mistaken for the reference cosmology used to generate the BBC simulations that estimates the $\mu_{\mathrm{bias}}$ term in equation~(\ref{eq:TrippMOD}).

Here, we explore replacing the fiducial cosmology (along with the fitted magnitude offsets in each bin) within the BBC fit with a spline interpolation of the SN magnitudes. To accomplish this, we modify the BBC procedure. Recall that within the current BBC procedure the Hubble residuals are minimized to a fiducial cosmology among 20 independent redshift bins, given a set of global nuisance parameters and 20 offsets in magnitude. Here, we instead minimise the Hubble residuals to a spline interpolation of the SN magnitudes, determined at each fitting step, where we used the weighted average redshift, $z_{\mathrm{avg}}$ and distance moduli, $\mu_{\mathrm{avg}}$ in 20 redshift bins as knots. 

We compare these two procedures by recreating a simplified BBC fitting procedure that attributes all of the intrinsic scatter to coherent variation at all epochs and wavelengths, $\sigma_{\rm int}$.\footnote{The baseline analysis (equation~\ref{eq:sig1}) instead uses $\sigma^2_{\rm floor}(z_i, c_i, M_{*,i})=\sigma^2_{\rm scat}(z_i, c_i, M_{*,i})+\sigma^2_{\rm gray}$ where $\sigma_{\rm scat}(z_i, c_i, M_{*,i})$ is determined from a model that describes intrinsic brightness fluctuations and $\sigma_{\rm gray}$ is determined after the BBC fitting process to bring the Hubble diagram reduced $\chi^2$ to $\sim 1$.} Further complexity is not required as the intrinsic scatter is incorporated into the uncertainties in the same way if we use a fiducial cosmology or a spline and we only need to test consistency between the two methods.\footnote{ Note the simplified fitting procedure means we will get slightly different values for $\alpha$ and $\beta$ than appear in Figs.~\ref{fig:Fid_test_cosmoabmean} and \ref{fig:100_combined}, but the values are not important, just whether they change between using a fiducial cosmology and a spline.}

Table~\ref{tab:spline_test} compares the fitted nuisance parameters using the same light curve sample when using two different fiducial cosmologies (see Table~\ref{tab:spline_test} for model parameters) and a spline that is determined at each fitting step. All parameters are consistent demonstrating the following. First, that the results from our simplified BBC fit are again insensitive to the choice of fiducial cosmology.  Second, that a spline is viable alternative to a fiducial cosmology and may reduce confusion as to the role of the fiducial cosmology in future pipelines. 

\begin{table}
\caption{BBC fitted nuisance parameters for three different fiducial cosmologies, showing the results are stable to the choice of fiducial cosmology or use of a spline (see Section~\ref{sec:isfidreq}).}
\label{tab:spline_test}
\renewcommand{\arraystretch}{1.3}
\begin{tabular}{cccc}
\hline \hline
& \multicolumn{3}{c}{Fiducial Cosmology}  \\[-1mm] 
Parameters &
\hspace{1mm}Flat-$\Lambda$CDM$^\dag$ &
\hspace{1mm}Flat-$w$CDM$^*$  &
\hspace{1mm}Spline  \vspace{-1mm} \\ \hline
$\sigma_{\mathrm{int}}$ & \hspace{1mm}$0.095^{+0.003}_{-0.004}$ &
\hspace{1mm}$0.098\pm 0.004$ &
\hspace{1mm}$0.099^{+0.003}_{-0.004}$\\
$\alpha$ & 
\hspace{1mm}$0.136\pm 0.004$ &
\hspace{1mm}$0.136^{+0.004}_{-0.005}$ &
\hspace{1mm}$0.137\pm0.004$\\
$\beta$ & 
\hspace{1mm}$3.008^{+0.040}_{-0.047}$ &
\hspace{1mm}$2.958^{+0.039}_{-0.048}$ &
\hspace{1mm}$2.978^{+0.040}_{-0.051}$ \\ \hline
\end{tabular}
{\raggedright $^{\dag}(H_0,\Omega_{\mathrm{m}}) =(70, 0.3)$\par}
{\raggedright $^{*}(H_0,\Omega_{\mathrm{m}}, w) =(60, 0.4, -0.8)$\par}
\end{table}

\section{Data}\label{sec:data}
Having established that the derivation of the DES-SN5YR Hubble diagram is robust to the choice of reference and fiducial cosmological models, we turn to using the Hubble diagram to derive constraints on a range of non-standard models which differ in their background expansion and are therefore sensitive to the DES-SN5YR data. To test the non-standard cosmology fitting code, we generated 25 simulations and ensured that fitted parameters of each model were consistent with the input cosmology. The input cosmology for these simulations used Flat-$\Lambda$CDM, for models that could reduce to Flat-$\Lambda$CDM for some values of their parameters. Otherwise, we used the model being tested as the input cosmology to generate the 25 realisations.

\subsection{The DES-SN5YR sample}
The DES-SN survey covers $\sim$27 deg$^2$ over 10 fields across the DES footprint \citep[see][]{DES_spec}. The survey, which ran for five years using the Dark Energy Camera \citep[DECam;][]{DECam}. DES detected over 30,000 SN candidates, from these 1635 were deemed SNe Ia-like and included in the DES-SN5YR Hubble diagram with 1499 photometrically classified as type Ia SNe using SuperNNova \citep{10.1093/mnras/stac1691, vincenzi24}. The DES-SN5YR sample includes publicly available low-$z$ SNe Ia from the Harvard-Smithsonian Center for Astrophysics, CfA3 \citep{Hicken_2009} and CfA4 \citep{Hicken_2012}, the Carnegie Supernova Project, CSP \citep{Krisciunas_2017_CSP} and the Foundation Supernova Survey \citep{2018_Foley_foundation}. These low-$z$ samples span a redshift range of 0.01 to 0.1. However, SNe Ia in the low-$z$ sample with redshifts $<0.025$ are excluded to minimise the impact of peculiar velocities. With this cut applied, the low-$z$ sample comprises 194 SNe, for a total of 1829 SNe in the DES-SN5YR sample; for more details see \citet{10.1093/mnras/stac1691, vincenzi24} and \citet{DESDATA24}.

\subsection{External probes}
 Our data must be interpretable in context of the parameters of the cosmological models that we test. In this work, many of these are defined as modifications to the background expansion and do not describe how the CMB or galaxy power spectrum may change. Additionally, we would like to be agnostic about the pre-recombination history, and in particular the size of the sound horizon $r_d$ or $r_*$. 
 
 Fortunately, as we describe below, we may still combine the DES-SN5YR cosmological constraints with measurements based on observations from the Cosmic Microwave Background (CMB) and Baryon Acoustic Oscillations (BAO) by the use of derived parameters with clear physical meaning. We do not use data from weak lensing surveys in this work.

\subsubsection{Cosmic Microwave Background}\label{cmbprior}
The CMB data may be expressed in terms of the `shift parameter' $R$ \citep{Bond1997}, defined in the literature as  
\begin{equation}
R=\sqrt{\Omega_m} S_k\left(\int_0^{z_{*}} \frac{d z}{E(z)}\right) \;\;,
\end{equation}
where $z_*$ is the redshift at the surface of last-scattering, $E(z)\equiv H(z)/H_0$ is the normalized redshift-dependent expansion rate and
\begin{equation}
S_k(x)= \begin{cases}\sin \left(\sqrt{-\Omega_k} x\right) / \sqrt{-\Omega_k} & \Omega_k<0, \\ x & \Omega_k=0, \\ \sinh \left(\sqrt{\Omega_k} x\right) / \sqrt{\Omega_k} & \Omega_k>0 .\end{cases}
\end{equation}
The physical meaning of $R$ in the context of non-standard cosmological models may be understood if the baryon density $\omega_b = \Omega_b h^2$ is fixed (for example by nucleosynthesis constraints). Although $R$ is sometimes interpreted as set by the location of the peaks and troughs of the CMB power spectrum (if the sound speed is fixed by $\omega_b$ and $T_{\rm CMB}$), this relies on the absence of additional energy components in the pre-combination era \citep[for example, early dark energy models as reviewed in][]{Poulin2023}. Alternatively, $R$ may also be understood as localised around the surface of last scattering in the following way. During recombination, photons stream out of over-densities and suppress power on small scales in a process known as Silk damping \citep{Silk1968}. Again at fixed $\omega_b$, successive spectral peaks are lower than their predecessors as the multipole $l$ increases, and the rate of suppression $C_l \propto \exp{-2(l/l_{\rm Silk})^2}$ \citep[see for example][]{Mukhanov2004} is proportional to the Hubble expansion rate at the time of last scattering. We may therefore define 
\begin{equation}
R' = \frac{H(z_*)D_{\rm M}(z_*)}{(1+z_*)^{3/2}}\;\;,
\end{equation}
where $D_M(z)$ is the transverse comoving distance defined as
\begin{equation}\label{eq:comov}
D_M(z)=\frac{c}{H_0} S_k\left(\int_0^{z} \frac{d z'}{E(z')}\right),
\end{equation}
We see that $R' \simeq R$ provided the universe is matter-dominated at the time of last scattering. It may be calculated that $R' \simeq 1.8 \times 10^{-3} l_{\rm Silk}$ where the prefactor is only sensitive to cosmological parameters by a factor of $(1+z_*)^{1/2}$ and in turn $z_*$ does not depend much on the cosmology. Hence $R'$, which is explicitly proportional to $H(z_*)$, connects $R$ to the Silk damping scale which we take as a safe assumption for the range of models we test. 

\cite{Chen_2019} converted the Planck 2018 \citep{2020_planck} TT,TE,EE + lowE measurements to a prior on $R$, finding $R=1.7502\pm0.0046$ for models assuming spatial flatness and $R=1.7429\pm0.0051$ for models that allow curvature. We use these priors in this work. We also note that \citet{Lemos23} remove late-time cosmology dependence from the CMB likelihoods by using flexible templates for late-ISW and CMB-lensing. We convert their baseline results \citep[Early-$\Lambda$CDM, see Table 1 of][]{Lemos23} into a constraint on the shift parameter and find $R=1.7442\pm0.0044$. Reassuringly, the central value falls between the constraints from \citet{Chen_2019}.

\subsubsection{Baryon Acoustic Oscillations}\label{baoprior}
Baryon acoustic oscillations represent a sharply-defined acoustic angular scale on the sky given by
\begin{align}
    \theta_d = \frac{r_d}{D_M(z_d)}\;\;,
\end{align}
where $D_M(z_d)$ is the transverse comoving distance to the drag epoch, and $r_d$ is the comoving sound horizon given by 
\begin{equation}
r_d=\int_{z_d}^{\infty} \frac{c_{s}(z)}{H(z)} d z
\end{equation}
and $c_{s}$ is the baryon sound speed, while $r_{*}$ and $\theta_*$ are defined in the same way using $z_*$.

BAO measurements are given as the ratio of $r_d$ to either the Hubble distance, $D_H(z)=c/H(z)$, transverse comoving distance, $D_M(z)$ or a combination of the two termed the dilation scale, $D_V(z) \equiv\left[z D_M^2(z) D_H(z)\right]^{1 / 3}$. To interpret these in terms of distances, $r_d$ is needed. However, in this work, we cancel the dependence on the sound horizon scale by using the ratio of the BAO distance with the distance to CMB as,
\begin{align}
\frac{D_M(z_*)}{D_{X_i}(z)} &= \frac{1}{\theta_*} \times \frac{r_d}{D_{X_i}(z)} \times \frac{r_*}{r_d} 
\end{align}
where $D_{X_i}=$ \{$D_{V}$, $D_{M}$, $D_{H}$\}, and we remind the reader that $D_{M}(z_*) = (c/H_0) R/\sqrt{\Omega_m}$. In this way, the data represents the ratio of the angular scales of the sound horizon on the surface of last scattering and at the effective redshift of the BAO. The cosmological dependence of $r_* / r_d $ may be neglected.

We use BAO data from the extended Baryon Oscillation Spectroscopic Survey \citep[eBOSS;][]{dawson16, alam21}, which is the cosmological survey within SDSS-IV \citep{blanton17}. Specifically, we use the BAO-only measurements from SDSS MGS \citep{ross15}, SDSS BOSS \citep{alam17}, SDSS eBOSS LRG \citep{bautista21}, SDSS eBOSS ELG \citep{demattia21}, SDSS eBOSS QSO \citep{hou21} and SDSS eBOSS Ly~$\alpha$ \citep{bourboux20}. We note that new BAO measurements from both DES \citep{desbao24} and the DESI collaboration \citep{desicollaboration2024desi} were released in the advanced stages of this work and motivates a follow-up analysis with with the inclusion of these data sets.

The covariance matrices provided by eBOSS\footnote{\url{https://svn.sdss.org/public/data/eboss/DR16cosmo/tags/v1_0_0/likelihoods/BAO-only/}} have been incorporated into this study with the use of the {\sc uncertainties} \citep{uncertainties} python package and the final measurements shown in Table~\ref{tab:BAOdata}. Note that although these measurements contain information from the CMB we will refer to these measurements as \bao from here on.

\begin{table}
\caption{Summary of the external constraints determined using measurements from eBOSS and Planck.}
\label{tab:BAOdata}
\renewcommand{\arraystretch}{1.3}
\begin{tabular}{cccc}
\hline \hline
\multicolumn{4}{c}{\textbf{\bao measurements$^*$}}  \\[-1mm] 
$z_{\text{eff}}$ & $D_M(z_*)/D_V(z)$ & $D_M(z_*)/D_M(z)$ & $D_M(z_*)/D_H(z)$ \\[-1mm] \hline
0.15 & $21.13 \pm 0.80$ & - & - \\
0.38 & - & $9.22 \pm 0.15$ & $3.78 \pm 0.11$\\
0.51 & - & $7.06 \pm 0.11$ & $4.23 \pm 0.11$\\
0.70 & - & $5.28 \pm 0.10$ & $4.88 \pm 0.14$\\
0.85 & $5.15 \pm 0.25$ & - & - \\
1.48 & - & $3.07 \pm 0.08$ & $7.12 \pm 0.30$\\
2.33 & - & $2.52 \pm 0.13$ & $10.58 \pm 0.34$\\
2.33 & - & $2.52 \pm 0.11$ & $10.42 \pm 0.36$\\ \hline \hline
\multicolumn{4}{c}{\textbf{\cmb measurements$^\dag$}} \\[-1mm] 
$z_*$ & $\Omega_{\rm k}$ & \multicolumn{2}{c}{$R$} \\[-1mm] \hline
1089.95 & $=0$ & \multicolumn{2}{c}{$1.7502\pm0.0046$} \\
1089.46 & $\neq 0$ & \multicolumn{2}{c}{$1.7429\pm0.0051$} \\
\hline
\end{tabular}
{\raggedright $^{*}$ The product of the BAO measurements with the CMB acoustic scale. \par}
{\raggedright $^{\dag}$ In this work we use the `shift parameter' $R$ that is related to the heights of the CMB acoustic peaks and depend on the line of sight distance to the sound horizon. \par}
\end{table}

\subsection{Constraining cosmological models}\label{constraining}
 In general, the parameters of an individual cosmological model are constrained by minimizing a $\chi^2$ likelihood given by
\begin{equation}\label{eq:chi2_none}
\tilde{\chi}^{2}=\vec{D}^{T}\left[C_{\text {stat+syst }}\right]^{-1} \vec{D}
\end{equation}
and for DES-SN5YR, $D_i = \mu_{\text{model},i} - \mu_i$ for the $i^{th}$ SN. However, the absolute magnitudes of SNe Ia are degenerate with $H_0$. For this analysis, no assumption on $H_0$ is presumed and instead $H_0$ is treated as a nuisance parameter that is analytically marginalised over by modifying equation~(\ref{eq:chi2_none}). The modified $\chi^2$ likelihood is given by \citep{Goliath_2001},
\begin{equation}\label{eq:chi2}
\chi^2_{\mathrm{SN}} =
\tilde{\chi}^{2}-\frac{B^{2}}{C}+\ln \left(\frac{C}{2 \pi}\right) 
\end{equation}
where
\begin{equation}
B =\sum_{i=1}^{n} \left(\left[C_{\text {stat+syst }}\right]^{-1} \cdot \vec{D} \right)_i
\end{equation}
and
\begin{equation}
C =\sum_{i=1}^{n} \sum_{j=1}^{n} \left[C_{\text {stat+syst }}\right]^{-1}_{ij} 
\end{equation}
and where we sum over all matrix elements, $i,j$. For the combined constraints we sum the $\chi^2$ likelihoods from all data sets as 
\begin{equation}\label{eq:chi2tot}
    \chi^2_{\mathrm{tot}} = \chi^2_{\mathrm{SN}} + \tilde{\chi}^2_{\mathrm{BAO-}\theta_*} + \tilde{\chi}^2_{\mathrm{CMB-}R}.
\end{equation}

{\sc cobaya}\footnote{\url{https://github.com/CobayaSampler/cobaya}} \citep{cobaya2, cobaya1}, a robust code for Bayesian analysis, was used to minimize equations~(\ref{eq:chi2})~and~(\ref{eq:chi2tot}). The convergence of MCMC chains was assessed in terms of a generalized version of the $R-1$ Gelman-Rubin statistic \citep{1992StaSc...7..457G}, which measures the variance between the means of the different chains in units of the covariance of the chains. For our work, we adopted a more stringent tolerance than {\sc cobaya}'s default value, namely $R-1=0.001$.

\section{Cosmological Models and Results}\label{sec:theory}
\begin{figure*}\centering
    \includegraphics[width=\linewidth]{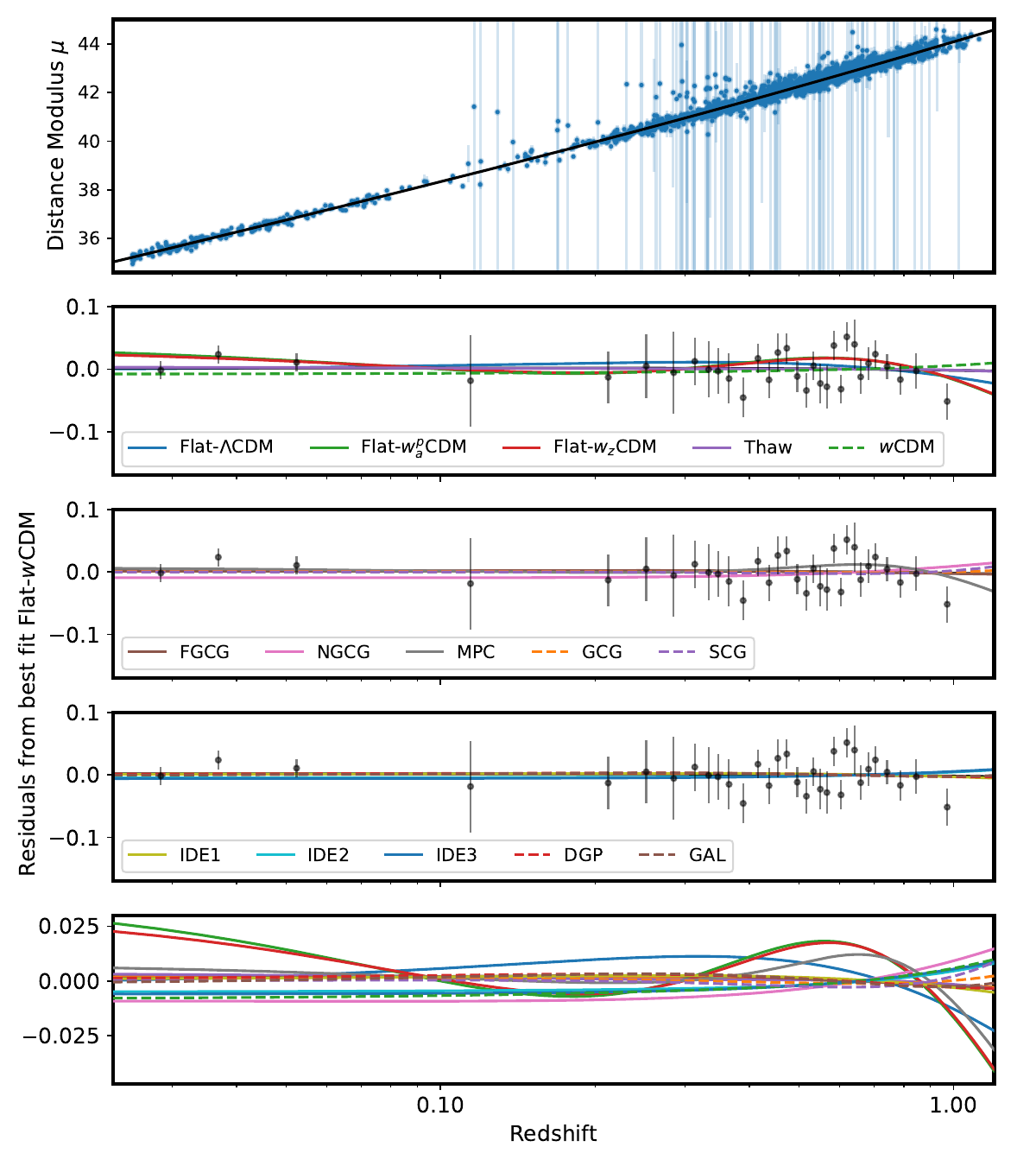}
    \caption{Upper Panel: Hubble diagram of DES-SN5YR with the overlaid best fit Flat-$w$CDM model. We also show the inflated distance uncertainties from likely contaminants. Four lower panels: The difference between the data and the best fit Flat-$w$CDM model from the DES-SN5YR alone. We also overplot the best fit for each model (we exclude the Timescape model as it was fit against a modified Hubble diagram). Spatially flat models are shown as solid lines and models that allow curvature are represented by dashed lines. }
    \label{fig:HD}
\end{figure*}

\KP~presents cosmological results for the standard cosmological model and simple variations such as allowing the dark energy equation of state to be other than $w=-1$ and/or vary with scalefactor. 
In this work, we extend on that analysis and present constraints on more exotic non-standard cosmological models. 

For each of the models we investigate, the same basic theory applies and the theoretical distance moduli can be calculated as,
\begin{equation}
    \mu(z) = 5 \log_{10}[\dl (z)] + 25.\label{eq:mu}
\end{equation}
$\dl(z)$  is the luminosity distance and follows the relation,
\begin{equation}\label{eq:dl}
\dl(z) = (1+z_{\rm obs})D_{\rm M}(z),
\end{equation}
where $z$ is the cosmological redshift and $z_{\rm obs}$ is the observed redshift.
However, the Friedmann equation (describing how the Hubble parameter changes with scalefactor or redshift) differs. 

In the following subsections, we briefly introduce each model and present the associated normalized Friedmann equation $E(z)$, used to determine $D_M(z)$ (equation~\ref{eq:comov}). We also present the associated parameter constraints using the DES-SN5YR sample alone and after combining the DES-SN5YR sample with the \cmb and \bao (summarised in Table~\ref{tab:results}). For all fits, we report the median of the marginalised posterior and cumulative $68.27$ per cent confidence interval. The best fit Hubble diagrams are shown in Fig.~\ref{fig:HD}.

\begin{table*}
\caption{Results for the cosmological models investigated in this work. These are the medians of the marginalised posterior with 68.27 per cent integrated uncertainties (`cumulative' option in ChainConsumer).}
\label{tab:results}
\renewcommand{\arraystretch}{1.3}
\begin{tabular}{lcccccc}
\hline \hline 
\textbf{Key Paper Results} & $\Omega_{\mathrm{m}}$ & $\Omega_{\Lambda}$ & $w_0$ & $w_{a}$  &  & \\[-1mm] \hline
\multicolumn{7}{l}{\textbf{DES-SN5YR }} \\
Flat-$\Lambda$CDM &  $0.352\pm0.017$ & - & - & - & - & -\\
$\Lambda$CDM &  $0.291^{+0.063}_{-0.065}$ & $0.55\pm 0.10$ & - & - & - & -\\
Flat-$w$CDM &  $0.264^{+0.074}_{-0.096}$ & - & $-0.80^{+0.14}_{-0.16}$ & - & -  & -  \\
Flat-$w_0 w_a$CDM &  $0.495^{+0.033}_{-0.043}$ & - & $-0.36^{+0.36}_{-0.30}$ & $-8.8^{+3.7}_{-4.5}$ & - & - \\ \hline
\textbf{Cosmography} & $q_0$ &$j_0$ &$s_0$ &  & & \\[-1mm] \hline 
\multicolumn{7}{l}{\textbf{DES-SN5YR }} \\
Third order &  $-0.362^{+0.067}_{-0.069}$ & $0.16^{+0.32}_{-0.29}$ & - &  &  &   \\
Fourth order &  $-0.06^{+0.11}_{-0.13}$ & $-2.43^{+0.92}_{-0.72}$ & $1.4^{+4.6}_{-3.3}$ &  &  & \\  \hline
\textbf{Parametric Models} & $\Omega_{\mathrm{m}}$ & $\Omega_{\rm de}$ & $w_0$ & $w_{z}$  & $w^p_0$ &  $w_{a}$ \\[-1mm]
 \hline
 \multicolumn{7}{l}{\textbf{DES-SN5YR}} \\
 $w$CDM  &  $0.262^{+0.068}_{-0.074}$ & $0.61^{+0.26}_{-0.25}$ & $-0.91^{+0.20}_{-0.43}$ & - & - & - \\
 Flat-$w_0 w_z$CDM &  $0.492^{+0.027}_{-0.038}$ & - & $-0.57\pm 0.23$ & $-6.0^{+2.5}_{-2.4}$ & - & - \\
 Flat-$w^{\rm p}_0 w_a$CDM where $z_p = 0.078$ &  $0.495^{+0.034}_{-0.045}$ & - & - & - & $-1.00^{+0.13}_{-0.14}$ & $-8.6^{+3.8}_{-4.5}$ \\
\multicolumn{7}{l}{\textbf{DES-SN5YR + \cmb + \bao}} \\
$w$CDM$^{\dag}$ & $0.320\pm0.007$ & $0.682\pm0.007$ & $-0.912\pm 0.029$ & - & - & -\\
Flat-$w_0 w_z$CDM$^{\dag}$ & $0.322\pm0.007$ & - & $-0.866^{+0.046}_{-0.042}$ & $-0.142^{+0.093}_{-0.123}$ & - & -  \\
Flat-$w^{\rm p}_0 w_a$CDM$^{\dag}$ where $z_p = 0.274$  &  $0.323\pm0.007$ & - & - & - & $-0.918\pm0.027$ & $-0.29^{+0.26}_{-0.28}$ \\ \hline
\textbf{Thawing Scaling Field Model} & $\Omega_{\mathrm{m}}$ & $w_0$ & $\alpha$ &  &  &   \\[-1mm] \hline
\multicolumn{7}{l}{\textbf{DES-SN5YR}} \\
Thaw & $0.306^{+0.041}_{-0.042}$ & $-0.83^{+0.12}_{-0.14}$ & $1.452^{+0.067}_{-0.068}$ & & & \\
\multicolumn{7}{l}{\textbf{DES-SN5YR + \cmb + \bao}} \\
Thaw & $0.323\pm0.007$ & $-0.867^{+0.041}_{-0.040}$ & $1.449^{+0.072}_{-0.065}$ & & & \\ \hline
\textbf{Chaplygin Gas} & $\Omega_{\mathrm{m}}$ & A & $\zeta$ & $w_0$  &  &   \\[-1mm] \hline
\multicolumn{7}{l}{\textbf{DES-SN5YR}} \\
SCG$^{*}$ &  $0.121\pm 0.035$ & $0.789^{+0.029}_{-0.027}$ & - & - & & \\
FGCG & $0.255^{+0.099}_{-0.133}$ & $0.600^{+0.049}_{-0.048}$ & $-0.33^{+0.33}_{-0.30}$ & - & & \\
GCG & $0.236^{+0.080}_{-0.124}$ & $0.65^{+0.15}_{-0.12}$ & $-0.01^{+1.09}_{-0.73}$ & - & & \\
NGCG & $0.278^{+0.095}_{-0.147}$ & $0.76^{+0.15}_{-0.27}$ & $0.03^{+1.15}_{-0.66}$ & $-0.78^{+0.16}_{-0.45}$ & & \\
\multicolumn{7}{l}{\textbf{DES-SN5YR + \cmb + \bao}} \\
SCG$^{*}$ & $0.376\pm0.009$ & $0.556\pm0.008$ & - & - & & \\
FGCG$^{\dag}$ & $0.322\pm0.007$ & $0.636^{+0.020}_{-0.019}$ & $-0.107^{+0.038}_{-0.035}$ & - & & \\ GCG$^{\dag}$ & $0.319\pm0.008$ & $0.634^{+0.021}_{-0.022}$ & $-0.120^{+0.042}_{-0.041}$ & - & & \\
NGCG & $0.323\pm0.007$ & $0.777^{+0.087}_{-0.125}$ &  $0.33^{+0.44}_{-0.40}$  & $-0.77^{+0.11}_{-0.20}$ & & \\ \hline 
\end{tabular}
{\par \raggedright $^{*}$ Cannot reduce to the cosmological constant for any set of parameters. \par}
{\raggedright $^{\dag}$ Best fits are $>2\sigma$ from the subset of parameters that reduce to the cosmological constant. \par}
\end{table*}

\begin{table*}
\contcaption{Results for the cosmological models investigated in this work. These are the medians of the marginalised posterior with 68.27 per cent integrated uncertainties (`cumulative' option in ChainConsumer).}
\renewcommand{\arraystretch}{1.3}
\begin{tabular}{lcccccc}
\hline \hline 
\textbf{Cardassian} & $\Omega_{\mathrm{m}}$ & $q$ & $n$ &  &  &   \\[-1mm] \hline
\multicolumn{7}{l}{\textbf{DES-SN5YR}} \\
MPC$^{\dag}$ & $0.467^{+0.032}_{-0.054}$ & $13.3^{+4.7}_{-6.5}$ & $0.464^{+0.034}_{-0.040}$ & & & \\
\multicolumn{7}{l}{\textbf{DES-SN5YR + \cmb + \bao}} \\
MPC & $0.322^{+0.007}_{-0.006}$ & $1.38^{+0.49}_{-0.42}$ & $0.25^{+0.12}_{-0.20}$ & & & \\  \hline
\textbf{Interacting Dark Energy} & $\Omega_{\mathrm{m}}$ & $w_0$ & $\varepsilon$ &  &  &   \\[-1mm] \hline
\multicolumn{7}{l}{\textbf{DES-SN5YR}} \\
IDE1 & $0.54^{+0.19}_{-0.32}$ & $-1.30^{+0.53}_{-0.91}$ & $0.46^{+0.90}_{-0.53}$ & & & \\
IDE2 & $0.31^{+0.22}_{-0.14}$ & $-0.85^{+0.17}_{-0.43}$ & $0.10^{+0.24}_{-0.36}$ & & & \\
IDE3 & $0.28^{+0.30}_{-0.21}$ & $-0.82^{+0.21}_{-0.60}$ & $ 0.12^{+0.86}_{-1.12}$ & & & \\
\multicolumn{7}{l}{\textbf{DES-SN5YR + \cmb + \bao}} \\
IDE1 & $0.53^{+0.18}_{-0.30}$ & $-1.38^{+0.55}_{-0.91}$ & $0.47^{+0.89}_{-0.54}$ & & & \\
IDE2$^{\dag}$ & $0.323\pm0.007$ & $-0.919\pm 0.032$ & $0.000\pm0.001$ & & & \\
IDE3 & $0.25^{+0.15}_{-0.10}$ & $-0.80^{+0.13}_{-0.26}$ & $-0.18^{+0.37}_{-0.28}$ & & & \\ \hline
\textbf{Modified  Gravity} & $\Omega_{\mathrm{m}}$ &  $\Omega_{\mathrm{k}}$ &  $\Omega_{\mathrm{rc}}$ &  $\Omega_{\mathrm{g}}$ &  &   \\[-1mm] \hline
\multicolumn{7}{l}{\textbf{DES-SN5YR}} \\
DGP$^{*}$ & $0.231^{+0.047}_{-0.051}$ & $0.03^{+0.18}_{-0.17}$ & $0.141^{+0.024}_{-0.025}$ & - & & \\
GAL$^{*}$ & $0.298^{+0.074}_{-0.073}$ &  $0.34\pm 0.15$ & - & $0.362^{+0.082}_{-0.078}$ & & \\
\multicolumn{7}{l}{\textbf{DES-SN5YR + \cmb + \bao}} \\
DGP$^{*}$ & $0.342\pm0.009$ & $0.014\pm0.003$ & $0.105\pm0.003$ & - & & \\
GAL$^{*}$ & $0.292\pm0.007$ & $-0.013\pm 0.004$ & - & $0.720\pm0.007$ & & \\  \hline
\textbf{Timescape} & $\Omega_{\mathrm{m}}^{\S}$ &  $f_{v0}$ &   &   &  &   \\[-1mm] \hline
\multicolumn{7}{l}{\textbf{DES-SN5YR$_{\mathrm{cut}}$}} \\
Timescape$^{*}$ & $0.292^{+0.043}_{-0.051}$ & $0.791^{+0.039}_{-0.034}$  & & & &  \\ 
Flat-$\Lambda$CDM & $0.362^{+0.019}_{-0.018}$ & -  \\ 
\multicolumn{7}{l}{\textbf{DES-SN5YR$_{\mathrm{cut}}$ + \baoperp}} \\ 
Timescape$^{*}$ &  $0.446^{+0.010}_{-0.009}$ & $0.665^{+0.008}_{-0.009}$   & & & & \\ 
Flat-$\Lambda$CDM &  $0.332^{+0.011}_{-0.010}$  & - & & & & \\
\hline
\end{tabular}
{\par \raggedright $^{*}$ Cannot reduce to the cosmological constant for any set of parameters. \par}
{\raggedright $^{\dag}$ Best fits are $>2\sigma$ from the subset of parameters that reduce to the cosmological constant. \par}
{\raggedright $^{\S}$ We convert the constraint on the void fraction to  the dressed matter density, which is related by $\Omega_{\mathrm{m} }=\frac{1}{2}\left(1-f_{\mathrm{v} 0}\right)\left(2+f_{\mathrm{v} 0}\right)$. \par}
\end{table*}

\subsection{Cosmography}\label{sec:cosmographic}
The cosmographic approach is a smooth Taylor expansion of the scale factor, $a$ that makes minimal assumptions about the underlying cosmological model, however retains the assumptions of homogeneity and isotropy \citep{Visser_2004, Zhang_2017, Macaulay_2019}. In cosmography, its useful to define the deceleration parameter,
\begin{equation}\label{eq:q_param}
    q = -\frac{1}{H^2}\frac{1}{a}\frac{d^2a}{dt^2}
\end{equation}
the jerk parameter,
\begin{equation}\label{eq:j_param}
    j = \frac{1}{H^3}\frac{1}{a}\frac{d^3a}{dt^3}
\end{equation}
and the snap parameter, 
\begin{equation}\label{eq:s_param}
    s = \frac{1}{H^4}\frac{1}{a}\frac{d^4a}{dt^4}
\end{equation}
where $q$ is directly related to the accelerated expansion of the universe and $j=1$ at all times for a spatially flat universe with a cosmological constant. Here, we Taylor expand the scale factor for a flat universe and take the series expansion to four terms, 
\begin{equation}\label{eq:cosmographic}
E(z)=\left[1+ \mathcal{C}_{1}z+ \mathcal{C}_{2}z^2+\mathcal{C}_{3}z^3+\mathcal{O}\left(z^4\right)\right]
\end{equation}
where $\mathcal{C}_{1} = \left(1+q_0\right)$, $\mathcal{C}_{2} = \frac{1}{2} \left(j_0-q_0^2\right)$, $\mathcal{C}_{3} = \frac{1}{6} \left( 3q_0^{2} + 3q_0^3 - 4q_0 j_0 - 3 j_0 - s_0\right)$ and $q_0$, $j_0$ and $s_0$ are the current epoch deceleration, jerk and snap parameters respectively.

\begin{figure*}\centering
    \includegraphics[height=0.4\textheight]{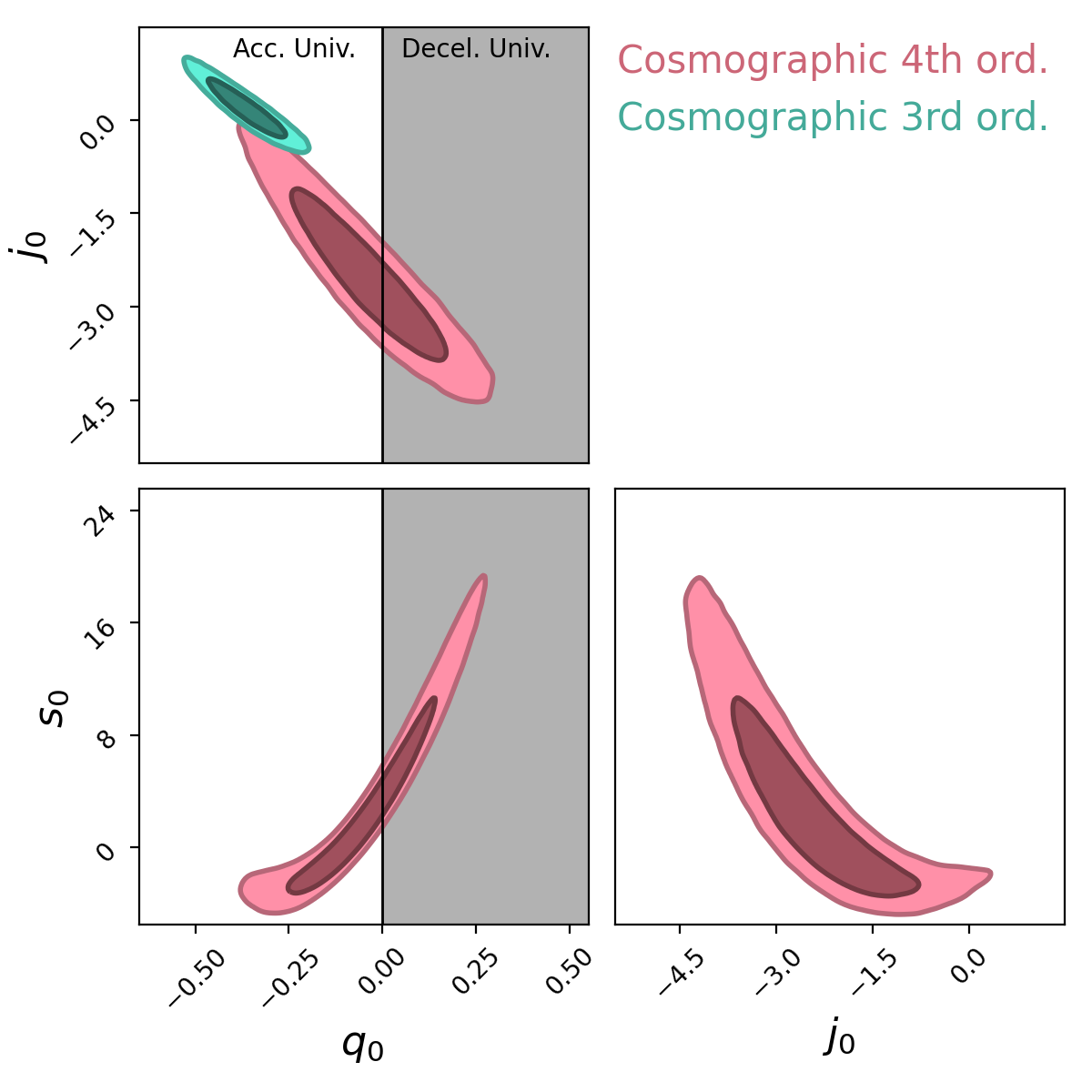}
    \caption{Constraints for the $3^{\rm rd}$ and $4^{\rm th}$ order cosmographic models (Section~\ref{sec:cosmographic}) from the DES-SN5YR data set only. The contours represent the 68.3 and 95.5 per cent confidence intervals.}
    \label{fig:contour_qjs}
\end{figure*}

We fit cosmographic expansion to third (equation~\ref{eq:cosmographic} excluding the $z^3$ term) and fourth order  (equation~\ref{eq:cosmographic}) with our constraints shown in Fig.~\ref{fig:contour_qjs}. For the third order fit we find $q_0=-0.362^{+0.067}_{-0.069}$ and evidence for an accelerating universe at $>5\sigma$. When we fit to fourth order, we find $q_0=-0.06^{+0.11}_{-0.13}$, which is consistent with zero however we note that the snap parameter is poorly constrained by the DES-SN5YR alone and find $s_0=1.4^{+4.6}_{-3.3}$. This result is analogous to the DES-SN5YR key paper results on the Flat-$w_0 w_a$CDM model who find a $w_0$ consistent with zero when $w_a$ is included in the fit. We also ensured that our fits were not over influenced by a particular redshift range and found consistent results after (a) removing low-$z$ data using the DES SNe alone and (b) removing high-$z$ SNe at $z>0.80$.

\begin{figure*}\centering
    \includegraphics[width=0.55\linewidth]{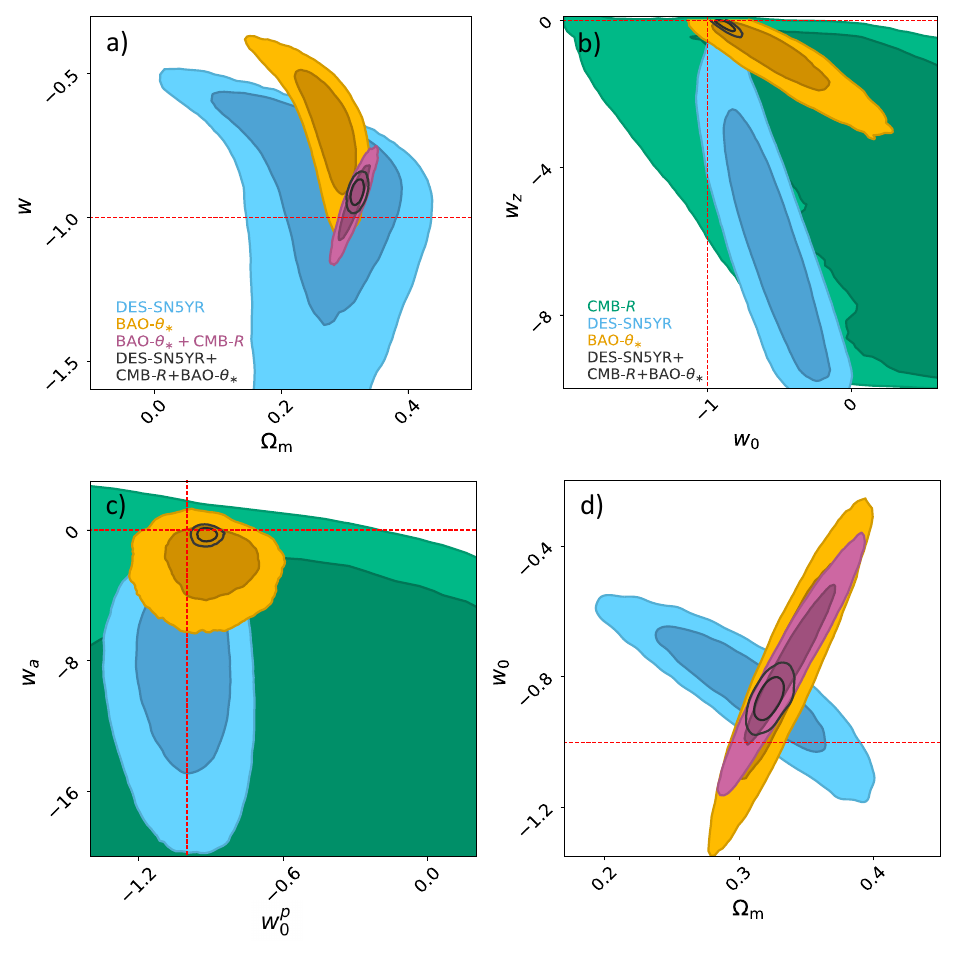}
    \caption{Parametric models, a) $w$CDM, b) Flat-$w_0 w_z$CDM, c) Flat-$w^{\rm p}_0 w_a$CDM (see Section~\ref{parametric_models}) and the d) thawing model (Section~\ref{sec:scalar_mod}): Constraints from the DES-SN5YR data set only (blue), a prior from the \cmb (green), \bao (orange), \cmb + \bao (purple) as well as the DES-SN5YR combined with both the \cmb and \bao priors (overlaid black contours). The contours represent the 68.3 and 95.5 per cent confidence intervals. The red dashed lines mark the parameters that recover a cosmological constant.}
    \label{fig:contour_qjw}
\end{figure*}

\subsection{Parametric models for the equation of state}\label{parametric_models}
The parametric models we consider here consider time varying dark energy with different functional forms of the dark energy equation of state, $w$. When all components have a constant equation of state, Friedmann's equation is simply
\begin{equation} \label{eq:wconst}
    E(z)^2=\sum_i\Omega_i a^{-3(1+w_i)}, 
\end{equation}
where the sum is over matter ($w_m=0$), curvature ($w_k=-1/3$), and dark energy with a constant equation of state ($w_{\rm de}={\rm constant}$), which could be a cosmological constant ($w_\Lambda=-1$).  Radiation ($w_r=1/3$) could also be included but is negligible for our redshift range. 
When testing dark energy with a time-varying equation of state one needs to make the substitution,
\begin{equation} \label{eq:wa}
    a^{-3(1+w_{\rm de})}\rightarrow \exp\left(3\int_a^1\frac{1+w(a)}{a}da \right).
\end{equation}

The simplest parametric model is where $w$ is generalised to an arbitrary constant while retaining spatial flatness (Flat-$w$CDM). This is the baseline cosmological model used within the DES-SN5YR analysis \KPp, who also test a flat model with a time varying dark energy in the form of $w(a)=w_0+w_a(1-a)$ (Flat-$w_0 w_a$CDM). While we do not refit these models here, we convert the constraints on Flat-$w_0 w_a$CDM using a linear variation of $w(a)$, which is anchored to a pivot redshift $z_p$ instead of $z=0$ (Flat-$w^{\rm p}_0 w_a$CDM), such that $w^{\rm p}(a)=w^{\rm p}_0+w^{\rm p}_a(a_{\rm p}-a)$ where $w_0^{\rm p}=w_0+w_a(1-a_{\rm p})$, $w_a^{\rm p}=w_a$, and $a_{\rm p}=1/(1+z_{\rm p})$. The pivot redshift corresponds to the redshift resulting in the tightest constraints on $w(a)$ \citep{Huterer_2001}. The expansion rate for the Flat-$w^{\rm p}_0 w_a$CDM model is given by
\begin{equation}
E(z)^2=\om a^{-3} + \Omega_{\rm de} a^{-3(1+w_0^{\rm p}+w_a^{\rm p} a_{\rm p} )}e^{-3w_a^{\rm p}(1-a)} 
\end{equation}
and in the case $z_p=0$, $a_{\rm p}=1$ the Flat-$w_0 w_a$CDM parameterisation is recovered.

We also test two other parameterisations. Firstly, the DES-SN5YR baseline model with spatial curvature as an additional free parameter ($w$CDM). Secondly, a model where $w(a)$ varies linearly in redshift instead of scalefactor (Flat-$w_0 w_z$CDM), such that $w(z) = w_0 + w_z z$ \citep{wz2002} and results in a Friedmann equation given by
\begin{equation} \label{eq:wz}
    E(z)^2=\Omega_{m}a^{-3}+ \Omega_{x}a^{-3(1+w_0-w_z)} e^{3w_z z}.
\end{equation}

Results for the parametric forms of the equation of state that we test within this work are summarised in Table~\ref{tab:results} and the associated contours are plotted in Fig.~\ref{fig:contour_qjw}.

Using the DES-SN5YR alone, the $w$CDM model is statistically consistent with a cosmological constant value of $w=-1$; however both Flat-$w_{z}$CDM and Flat-$w^{\rm p }_{a}$CDM favour a time-varying component to $w$ that increases with time. We note that the Flat-$w_0w_a$CDM model was constrained in \KP~finding $(w_0, w_a)=(-0.36^{+0.36}_{-0.30}, -8.8^{+3.7}_{-4.5}$) using DES-SN5YR alone. Here, we refit and convert these results to the equation of state at the pivot redshift and find $(w^{p}_0, w_a, z_p) = (-1.00^{+0.13}_{-0.14}, -8.6^{+3.8}_{-4.5}, 0.078)$.

When we combine DES-SN5YR with the \cmb and \bao our results are still consistent with a time-varying component to $w$ that increases with time with the best fit $w_{z}$ and $w_a$ (we find $z_p=0.274$ for the pivot redshift) both remaining $>1\sigma$ from a static $w$.

Interestingly, with the combined data sets, all parametric forms of the dark energy equation of state result in a best fit $w>1\sigma$ from a cosmological constant and all favour a $w>-1$. 

\subsection{``Thawing" scalar field models}
\label{sec:scalar_mod}

Light scalar fields provide a dynamical model for evolving dark energy inspired by scalar field models for primordial inflation. In the simplest incarnation of these models, the true vacuum energy density (or cosmological constant) of the universe is assumed to be zero, and dark energy is a transient phenomenon arising from the fact that a classically evolving scalar field $\phi$ with effective mass $m_\phi \la H_0$ has not yet have reached its ground state. In most particle physics models, light scalars are not technically natural, so it is conventional to consider models in which the small scalar mass is protected by a weakly broken shift symmetry, as is the case for the pseudo-Nambu-Goldstone boson (PNGB) model introduced by \cite{1995_Frieman}. 

Assuming the canonical Lagrangian for a scalar field, ${\cal{L}}= (1/2)g^{\mu\nu}\partial_\mu \phi \partial_\nu \phi -V(\phi)$, neglecting spatial perturbations the equation of motion of the field in an expanding universe is given by
\begin{equation} \label{eq:phiEOM}
\ddot{\phi}+3H\dot{\phi}+\frac{dV}{d\phi}=0~,
\end{equation}
where the expansion rate is given by
\begin{equation}
H^2 = \frac{8\pi}{3M_{Pl}^2}\left(\rho_m +\rho_\phi\right)~,
\end{equation}
$M_{Pl}=G^{-1/2}$ is the Planck mass, and the energy density of the field is 
\begin{equation}
\rho_\phi = \frac{1}{2}\dot{\phi}^2+V(\phi)~.
\end{equation}
The time-evolution of $\rho_\phi$ is determined by $H$ and by the equation of state parameter, $w_\phi=p_\phi/\rho_\phi$, where the scalar field pressure is 
\begin{equation}
p_\phi=\frac{1}{2}\dot{\phi}^2-V(\phi)~.
\end{equation}
For a given form of the potential $V(\phi)$ and initial value of the scalar field, $\phi(t_i) \equiv \phi_i$ at some early time $t_i \ll t_0$, this dynamical system can be solved to obtain $\phi(t)$ and thus the expansion history (assuming spatial flatness)
\begin{equation}
E(z)^2=\Omega_m a^{-3}+ \frac{\rho_\phi}{\rho_{crit}}~,
\end{equation}
where $\rho_{crit}=3M_{Pl}^2H_0^2/8\pi$. 

For ``thawing" scalar field models (the thawing/freezing nomenclature is from \cite{2005_Caldwell}), which include standard potentials of the form $V=(1/2)m_\phi^2 \phi^2 + \lambda \phi^4$ (with $\lambda>0$), the PNGB model $V(\phi)=m^2 f^2(1-\cos(\phi/f))$, and polynomials $V(\phi)=\sum_{i=1}^{n}a_i\phi^i$ with $a_i \geq 0$, at early times the driving term $dV/d\phi$ in equation~(\ref{eq:phiEOM}) is subdominant compared to the Hubble-damping term $3H\dot{\phi}$. In this limit, the field is effectively frozen at its initial value $\phi_i$, hence $\dot{\phi}(t_i)=0$, $\rho_\phi(t_i)=V(\phi_i)$, and $w_\phi(t_i)=-1$. Once the expansion rate drops below the curvature of the potential, $H \la \sqrt{|d^2V/d\phi^2|}$, the field begins to roll down the potential, develops non-negligible kinetic energy, and $w_\phi$ grows from $-1$. The parameters of $V(\phi)$ and the value of $\phi_i$ jointly determine $w_\phi(t)$ and the current scalar energy density, $\Omega_\phi=\rho_\phi(t_0)/\rho_{crit}$. 

For example, for a free, massive scalar with $V=(1/2)m_\phi^2 \phi^2$ the condition $\Omega_m=0.3$ implies $(m_\phi/H_0)(\phi(t_0)/M_{Pl}) \simeq 0.4$ in the limit where $\dot{\phi}^2 \ll V(\phi)$. For $m_\phi/H_0 \ga 1$ $(\la 1)$ the field begins rolling before the present epoch (or not) and the present value of the equation of state parameter, $w_0 \equiv w_\phi(t_0)$, can be measurably above $-1$ (or not), $w_0 \simeq -1+(1/7)(m_\phi/H_0)^2$.

While there have been a variety of approximate solutions and fits to late-time scalar field evolution (e.g., \cite{2008_Dutta, 2008_dePutter, 2009_Chiba}), numerical experiments show that the redshift-evolution of $w_\phi$ for thawing models is very well approximated by 
\begin{equation} \label{eq:walpha}
w_\phi(z)=-1+(1+w_0)e^{-\alpha z}~,
\end{equation}
where the value of $\alpha$ is only very weakly dependent on $w_0$ and on the form of $V(\phi)$ and is generally in the narrow range $\alpha=1.35-1.55$. As a consequence, these models are characterized by a quasi-one-dimensional parameter space that can be taken to be $w_0$ (with $\alpha=1.45\pm 0.1$). This approximation holds if the effective scalar mass $m_\phi$ is not large compared to $H_0$ (otherwise, the field will begin oscillating around the minimum of its potential by the present epoch.)

In Fig.~\ref{fig:contour_qjw}, we show constraints on $w_0$ and $\Omega_m$ marginalized over the narrow thawing-model prior on $\alpha$. For DES-SN5YR alone, we find $\Omega_m=0.306^{+0.041}_{-0.042}$ and $w_0=-0.83^{+0.12}_{-0.14}$; including CMB and BAO measurements, the resulting constraints are $\Omega_m=0.323\pm 0.007$ and $w_0=-0.867^{+0.041}_{-0.040}$, i.e., a $3\sigma$ deviation from $w_0=-1$. As shown in Table~\ref{tab:model_comp}, for the combined data sets the thawing model is moderately preferred over $\Lambda$CDM based on the AIC. 

The current data provide no meaningful constraint on the parameter $\alpha$ that determines the speed with which $w_\phi$ grows from its asymptotic value of $-1$. 
That is, if we widen the theory prior on $\alpha$ to allow values $\alpha \gg 1$, the best-fitting values are very large, with very large uncertainties. Note that for $\alpha \gg 1$, $w(z)=-1$ down to very low redshift $z \ll 1$, so cosmic distances vs. redshift 
should be indistinguishable from those in $\Lambda$CDM.

\subsection{Chaplygin gas models}\label{sec:chap_mod}

Chaplygin gas models deviate from $\Lambda$CDM by invoking an exotic background fluid with an equation of state $p= -A \rho^{-\zeta}$ \citep{2001kamen, 2002GCG, 2004Gchap} where $A$ is a positive constant. Chaplygin gas models represent pressureless dark matter in the early universe and dark energy in recent times and therefore may also be able to unify dark matter and dark energy \citep{2002chap}.

The simplest form of Chaplygin gas, which was introduced by \cite{2001kamen}, has an equation of state $p\propto \rho^{-1}$ ($\zeta = 1$). This model is referred to as the Standard Chaplygin Gas (SCG) model with a Friedmann equation given by 
\begin{equation}\label{eq:SCG}
E(z)^2=\frac{\Omega_{k}}{a^{2}}+\left(1-\Omega_{k}\right) \sqrt{A+\frac{(1-A)}{a^{6}}}.
\end{equation}
SCG has been shown to be inconsistent with other data sets \citep{2003chapgone, 2004chapgone, Davis_2007} however will be re-tested within this work. 

Generalised Chaplygin Gas (GCG), which maintains $\zeta$ as a free parameter, results in a Friedmann equation given by
\begin{equation} \label{eq:GCG}
E(z)^2=\frac{\Omega_k}{a^{2}}+\left(1-\Omega_k\right)\left(A+\frac{(1-A)}{a^{3(1+\zeta)}}\right)^{\frac{1}{1+\zeta}}
\end{equation}
and reduces to $\Lambda$CDM for $\zeta = 0$ and $\Omega_m = (1-\Omega_k)(1-A)$. 

We note that as $\Lambda$CDM is recovered for $\zeta = 0$, the SCG model (which has $\zeta=1$) cannot reduce to $\Lambda$CDM for any parameter choice. As a result it may not be surprising that, in contrast to the SCG model, 
the GCG model has been shown to be consistent with the previous data combinations \citep{Davis_2007, 2008GCG, Sollerman_2009, 2010GCG, Zhai_2017} consisting of the ESSENCE, SDSS-II, Constitution and Pantheon SN data sets \citep{Wood_Vasey_2007, Sako_2008, Hicken_2009, 2018Panth}. 

\cite{2008GCG} suggest that GCG can be thought of as an interacting form of $\Lambda$CDM. The analogous interacting form of $w$CDM was proposed by \cite{2006NGCG} termed New Generalised Chaplygin Gas (NGCG). The Friedmann equation for the spatially flat NGCG model is given by
\begin{equation}
E(z)^2=a^{-3}\left[1-A\left(1-a^{-3 w(1+\zeta)}\right)\right]^{\frac{1}{1+\zeta}}
\end{equation}
and can be reduced to $w$CDM for $\zeta = 0$.

\begin{figure*}\centering
    \includegraphics[height=0.4\textheight]{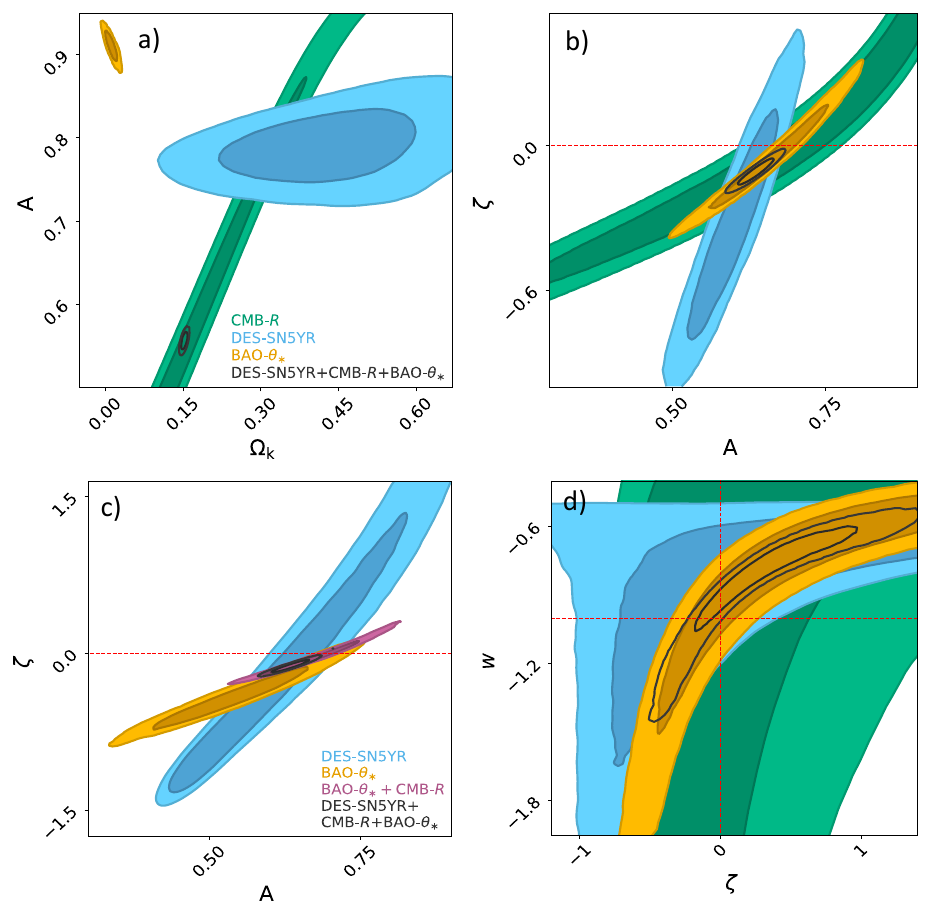}
    \caption{Chaplygin Gas models, a) SCG, b) FGCG, c) GCG and d) NGCG (Section~\ref{sec:chap_mod}): Constraints from the DES-SN5YR data set only (blue), a prior from the \cmb (green), \bao (orange), \cmb + \bao (purple) as well as the DES-SN5YR combined with both the \cmb and \bao priors (overlaid black contours). The contours represent the 68.3 and 95.5 per cent confidence intervals. The red dashed lines mark the parameters that recover a cosmological constant.}
    \label{fig:contour_CG}
\end{figure*}

In Fig.~\ref{fig:contour_CG} we present the contours for the Chaplygin gas models we investigate in this work. Constrained by the DES-SN5YR alone, the SCG model provides the lowest central value for the matter density of all models tested within this work at $\Omega_{\rm m}=0.121\pm0.035$. We note that this is due to the model favouring a high curvature, equivalent to $\Omega_{\rm k}=0.43\pm 0.12$. When combined with external priors, the SCG model is unable to simultaneously fit the different data sets (see Fig.~\ref{fig:contour_CG}a), which show extremely strong disagreement in the best fit parameters and highlighted by the poor Akaike Information Criterion (AIC) result of $\Delta$AIC$=276.9$ relative to Flat-$\Lambda$CDM (see Section~\ref{sec:model_comp} and Table ~\ref{tab:model_comp}).

Using the DES-SN5YR alone, the remaining Chaplygin Gas models FGCG, GCG and NGCG are consistent within $1\sigma$ ($\zeta=0$ and $w=-1$ for NGCG) of a cosmological constant. When combined with the CMB-$R$ and BAO-$\theta_*$ both the FGCG and GCG models find $\zeta>1\sigma$ from a cosmological constant. For the NGCG model, the best fit $\zeta$ is consistent with a cosmological constant, however favouring $w>-1$.

\subsection{Cardassian models}\label{sec:cardassian}

Cardassian models, first proposed by \cite{2002card} deviate from $\Lambda$CDM with the following modification to the Friedmann\-/Lema\'itre\-/Robertson\-/Walker metric (FLRW) equation,
\begin{equation}
    \label{eq:card}
    H^2 =A\rho+B\rho^n
\end{equation}
where the usual FLRW equation is recovered for $B=0$. Cardassian models invoke no vacuum energy ($\Lambda = 0$), instead the additional term in equation~(\ref{eq:card}) ($B\rho^n$) is initially negligible and only begins to dominate in recent times. Once the second term dominates, it causes the universe to accelerate. Therefore, with this modification, pure matter (or radiation) alone can drive an accelerated expansion. Some motivations for the addition of this term have been suggested and include self-interaction of dark matter \citep{gondolo2002accelerating}, as well as the embedding of our observable three-dimensional brane in a higher-dimensional universe \citep{2000brain}. The original power-law Cardassian model results in a Friedmann equation of the same functional form as that of $w$CDM where $w = n-1$ and therefore does not need to be tested separately. \cite{2003wang} generalises this model by introducing an additional free parameter $q > 0$. This model is called Modified Polytropic Cardassian (MPC) expansion which follows,

\begin{equation}
\label{eq:cardassian}
E(z)^2=\frac{\Omega_{m}}{a^{3}}\left(1+\frac{\left(\Omega_{m}^{-q}-1\right)}{a^{3 q(n-1)}}\right)^{\frac{1}{q}}
\end{equation}
and collapses to Flat-$w$CDM for $q=1$ where $w = n-1$. 

Our constraints in the $n-q$ plane for MPC expansion are shown in Fig.~\ref{fig:contour_MPCIDE}a. We find $q=13.3^{+4.7}_{-6.5}$ using DES-SN5YR alone, inconsistent with $q=1$ by $\sim 2\sigma$. This result is inconsistent with previous analyses by \citet{Zhai_2017} and \citet{Maga_a_2018} however these analyses both include constraints from probes other than SN. 
Our results are consistent with these previous analyses and $q=1$ when we supplement the DES-SN5YR data with external probes, we find $q=1.38^{+0.49}_{-0.42}$.

\subsection{Interacting dark energy \& dark matter}\label{sec:IDE}
In typical cosmological models, dark matter and dark energy are assumed to evolve independently. However, dark energy and dark matter provide the largest contribution to the energy budget of the universe so it is worth investigating if these components can interact. Interacting dark energy \& dark matter (IDE) models are therefore those which allow for this interaction \citep{1987_Freese} and are desirable as they allow solutions with a constant dark energy to matter ratio, solving the coincidence problem.

\begin{figure*}\centering
    \includegraphics[height=0.4\textheight]{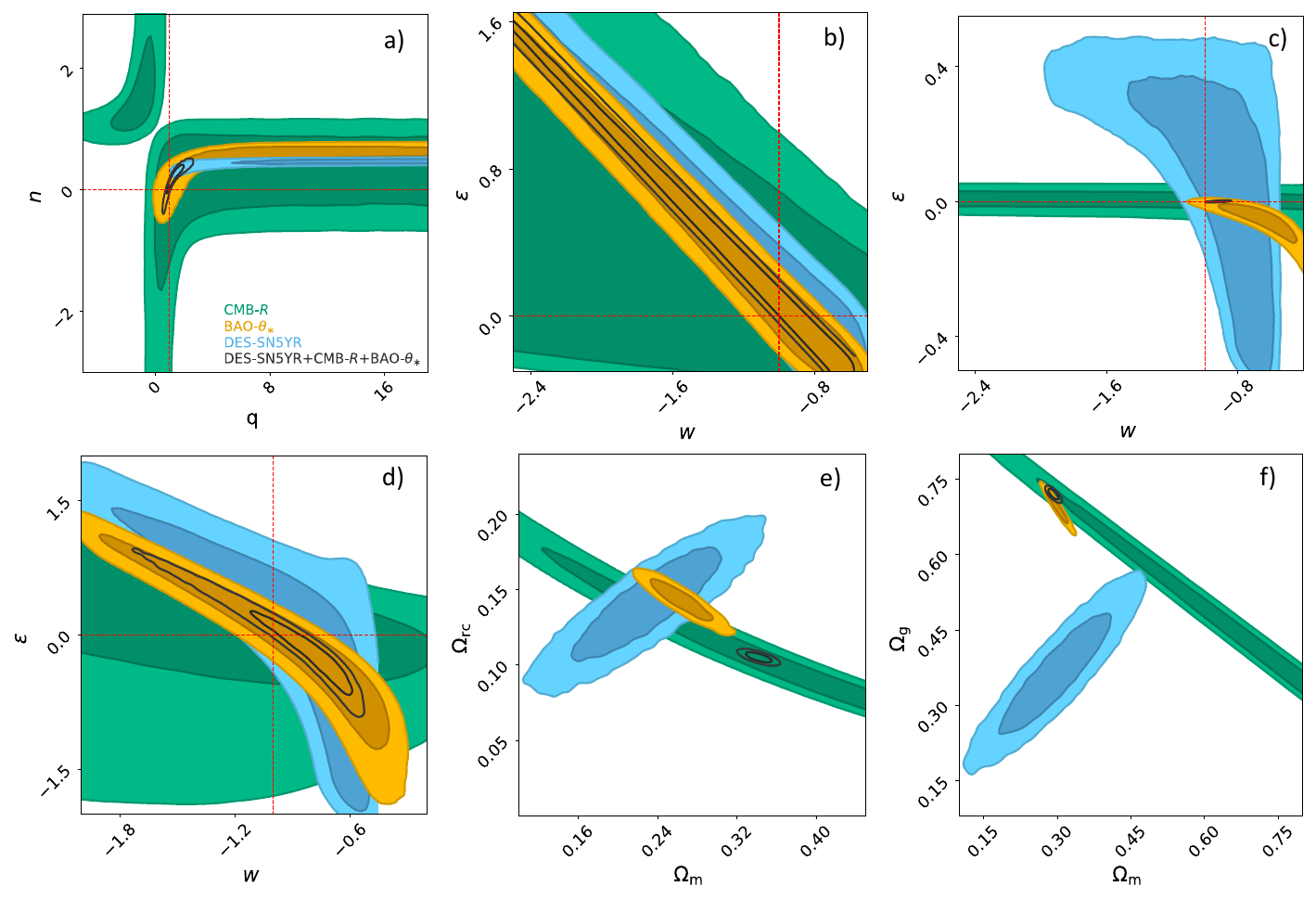}
    \caption{Same as Fig.~\ref{fig:contour_CG} but for the a) MPC model (Section~\ref{sec:cardassian}), the three IDE models (Section~\ref{sec:IDE}): b) IDE1, c) IDE2 and d) IDE3, as well as the e) DGP model and f) GAL model (Section~\ref{sec:GR_mod}).}
    \label{fig:contour_MPCIDE}
\end{figure*}

In this paper, we consider a popular subset    \citep{Barnes_Francis_Lewis_Linder_2005, 2007_Guo_IDE,2009_Miao_IDE,2010_He_IDE,2014_Li_IDE,2016_Hu_IDE_SN, IDE2016, IDE2019} of IDE models where the total energy density of dark energy and dark matter is conserved however the particular densities evolve as,

\begin{equation}
\begin{array}{l}
\dot{\rho}_{\rm m}+3 H \rho_{\rm m}=Q \\
\dot{\rho}_{x}+3 H\left(1+w\right) \rho_{x}=-Q
\end{array}
\label{eq:DEs}
\end{equation}
where ${\rho}_{\rm m}$ and ${\rho}_{x}$ represent the density of matter and dark energy respectively, $w$ is the dark energy equation of state and $Q$ is the interaction kernel which indicates the rate of energy transfer between the two components. 

We investigate three spatially flat IDE models where $Q$ has the general form $Q = H\varepsilon f(\rho_x, \rho_{\rm m})$, the function $f(\rho_x, \rho_{\rm m})$ specifies a particular IDE model and $\varepsilon$ is the coupling parameter between the dark components. The sign of $\varepsilon$ describes the energy flow between the interacting components where $\varepsilon < 0$ corresponds to a flow of energy from dark matter to dark energy. The paramaterizations of $Q$ and the respective Friedmann equations are:
\begin{enumerate}
  \item {\rm IDE1:} $Q = 3H\varepsilon \rho_{x}$ 
\begin{equation} \label{eq:IDE1}
    E(z)^2=\Omega_{\mathrm{m}}a^{-3} +\Omega_{x}\biggr[\frac{\varepsilon~ a^{-3}}{w + \varepsilon} + \frac{w~ a^{-3\left(1+w+\varepsilon\right)}}{w + \varepsilon}\biggr]
\end{equation} 
  \item {\rm IDE2:} $Q = 3H\varepsilon \rho_{\rm m}$ 
\begin{equation} \label{eq:IDE2}
    E(z)^2=\frac{\Omega_{x}}{a^{3\left(1+w\right)}} \scalemath{1}{+} \Omega_{\mathrm{m}}\scalemath{1}{\biggr[{ \frac{\varepsilon  ~a^{-3\left(1+w\right)}}{w + \varepsilon}\scalemath{1}{+}\frac{w~a^{-3\left(1-\varepsilon\right)}}{w + \varepsilon}} \biggr]}
\end{equation}
  \item {\rm IDE3:} $Q = 3H \varepsilon \frac{\rho_{\rm m} \rho_x}{\rho_{\rm m} + \rho_x}$
\end{enumerate}  
\begin{equation} \label{eq:IDE3}
    E(z)^2=  \frac{\Omega_{\mathrm{m}}~ C(a)}{a^{3}} + \frac{ \Omega_{x}~C(a)}{a^{3\left(1+w + \varepsilon\right)}}
\end{equation}
where
\begin{equation}
C(a) = \left[\frac{\Omega_{\mathrm{m}}}{\Omega_{\mathrm{m}} + \Omega_{x}} + \frac{\Omega_{x}}{\Omega_{\mathrm{m}} + \Omega_{x}}a^{-3\left(w+\varepsilon\right)}\right]^{\frac{-\epsilon}{ \epsilon+w}}. \nonumber 
\end{equation} 
The IDE models in equations~(\ref{eq:IDE1}),~(\ref{eq:IDE2})~and~(\ref{eq:IDE3}) will be referred to respectively as IDE1, IDE2 and IDE3 throughout this work. The results for the three IDE models we test are summarised in Table~\ref{tab:results} and the contours are shown in Fig.~\ref{fig:contour_MPCIDE}.

Using DES-SN5YR alone and after combining the DES-SN5YR with priors from the \cmb and \bao all of the IDE models tested are consistent within $1\sigma$ of no interaction between the dark components, $\varepsilon=0$ and $w=-1$. 

We also note that the \cmb puts a stringent constraint on the interaction for the IDE2 model, where find $\varepsilon = 0.000\pm0.001$. The tightness of the constraint on $\varepsilon$ is expected and in agreement with previous works \citep{2007_Guo_IDE,IDE2016}. This is due to the \cmb data not allowing a large deviation from the standard matter-dominated epoch along with the second term in equation~(\ref{eq:IDE2}).

\subsection{Modified gravity}\label{sec:GR_mod}
Dvali-Gabadadze-Porrati (DGP) brane world models first introduced by \cite{Dvali_2000} arise from a mechanism where the observed 4D gravity is embedded on a brane in 5D Minkowski space. As a result, locally the gravitational potential propagates in 4 dimensions reducing to General Relativity. However, at large distances the gravitational potential propagates in 5D or `leaks out into the bulk' deviating from General Relativity and causing accelerated expansion. Two branches of cosmological solutions in the DGP model have distinct properties. The solution examined in this work is the so-called self-accelerating branch where the late-time acceleration of the universe occurs without the need of a cosmological constant \citep{2001_DGP_BRANCH} and is described by
\begin{equation}\label{eq:DGP}
E(z)^2 =\frac{\Omega_k}{a^{2}}+\left(\sqrt{\frac{\Omega_{m}}{a^{3}}+\Omega_{r_{c}}}+\sqrt{\Omega_{r_{c}}}\right)^{2}
\end{equation}
where $\Omega_{m} =1-\Omega_k-2 \sqrt{\Omega_{r_{c}}} \sqrt{1-\Omega_k}$ and the length scale for which the `leaking' takes place is $r_{c}$ and $\Omega_{r_{c}} = 1/4 r^2_{c} H_{0}^{2}$. Therefore, the Flat-DGP and DGP models have the same number of free parameters as Flat-$\Lambda$CDM and $\Lambda$CDM respectively.

Inspired by the DGP model, \cite{2009_Galileon,2009_Galileon2} introduced Galileon cosmology, which is a scalar field class of models that are invariant under a shift symmetry in field space. Importantly, the Galileon scalar has no effect on the expansion rate during early times due to a natural screening mechanism, the Vainshtein effect in which non-linear effects can effectively decouple the scalar field from gravity \citep{2011_Gal_Felice}.  In late times, there exists a tracker solution (GAL) that is stable and self-accelerating with a very negative equation of state $w<-1$. The Friedmann equation for the GAL model has the same number of free parameters as $\Lambda$CDM and is given by
\begin{equation}\label{eq:GAL}
E(z)^2 =\frac{\Omega_k}{2a^{2}}  +\frac{\Omega_{m}}{2a^{3}}  + \sqrt{\Omega_{g}+\frac{1}{4a^{4}}\left[\frac{\Omega_{m}}{a}+\Omega_k\right]^{2}} 
\end{equation}
where $\Omega_{g}=1-\Omega_m - \Omega_k$. 
Both the DGP and GAL models provide a good fit to DES-SN5YR alone. However, when we include external probes, our results (summarised in Table~\ref{tab:results}) are in agreement with previous works \citep{2009_DGPOUT1,Li_2011, 2016_DECompare_planck, Zhai_2017, 2018_GALOUT} that show the DGP and GAL models to be inconsistent with multiple data sets, as seen in Figs.~\ref{fig:contour_MPCIDE}e~\&~\ref{fig:contour_MPCIDE}f.

\subsection{Timescape cosmology}\label{sec:timescape}
So far, the models examined all seek to explain the observed acceleration of the universe, assuming a FLRW geometry which is exactly homogeneous and isotropic. However, the local Universe is far from homogeneous and possesses a cosmic web of structures dominated in volume by voids. Timescape cosmology \citep{Wiltshire_2007a, Wiltshire_2007b, Wiltshire_2009} discards the approximation of a FLRW universe and instead considers a Buchert average \citep{Buchert_2000} over spatially flat wall regions and negatively curved voids. While the Buchert formalism has been investigated in other works, Timescape cosmology also accounts for a geometry difference between the Buchert average and an observer in a gravitationally bound system within the wall regions, for a universe dominated by voids. \cite{Wiltshire_2008} shows that this two-scale model results in a difference in clock rates that accumulates over cosmic time. In this work we use the Timescape tracker solution where the luminosity distance is calculated as,
\begin{equation}\label{eq:TS_dl}
\begin{aligned}
d_{\mathrm{L}} =(1+z)^2 t^{2 / 3}\left(\mathcal{F}\left(t_0\right)-\mathcal{F}(t)\right)
\end{aligned}
\end{equation}
where 
\begin{equation}\label{eq:TS_ft}
\begin{aligned}
\mathcal{F}(t) \equiv 2 t^{1 / 3} & +\frac{b^{1 / 3}}{6} \ln \left(\frac{\left(t^{1 / 3}+b^{1 / 3}\right)^2}{t^{2 / 3}-b^{1 / 3} t^{1 / 3}+b^{2 / 3}}\right) \\
& +\frac{b^{1 / 3}}{\sqrt{3}} \tan ^{-1}\left(\frac{2 t^{1 / 3}-b^{1 / 3}}{\sqrt{3} b^{1 / 3}}\right),
\end{aligned}
\end{equation}
$t$ is defined implicitly in terms of the redshift by
\begin{equation}\label{eq:TS_z1}
z+1=\frac{2^{4 / 3} t^{1 / 3}(t+b)}{f_{\mathrm{v} 0}^{1 / 3} \bar{H}_0 t(2 t+3 b)^{4 / 3}}
\end{equation}
and
\begin{equation}\label{eq:TS_b}
b \equiv 2\left(1-f_{\mathrm{v} 0}\right)\left(2+f_{\mathrm{v} 0}\right) /\left(9 f_{\mathrm{v} 0} \bar{H}_0\right).    
\end{equation}
Note that $f_{\mathrm{v} 0}$ is the current epoch void fraction and the only free parameter of the Timescape model (as we treat $H_0$ as a nuisance parameter in this work), which is related to the $\textit{dressed}$\footnote{The dressed parameters are defined such that they take numerical values similar to those of cosmological parameters within FLRW models.} matter density parameter by
\begin{equation}\label{eq:TS_om}
\Omega_{\mathrm{m}}=\frac{1}{2}\left(1-f_{\mathrm{v} 0}\right)\left(2+f_{\mathrm{v} 0}\right).
\end{equation}
The time, $t$ and Hubble parameter, $\bar{H}_0$ in equations~(\ref{eq:TS_dl}),~(\ref{eq:TS_ft}),~(\ref{eq:TS_z1})~and~(\ref{eq:TS_b}) are the volume averaged values, which are related to values we observe in a wall region by
\begin{equation}
H_0=\frac{\left(4 f_{\mathrm{v} 0}^2+f_{\mathrm{v} 0}+4\right)\bar{H}_0}{2\left(2+f_{\mathrm{v} 0}\right)}
\end{equation}
and
\begin{equation}
\tau=\frac{2}{3} t+\frac{4 \Omega_{\mathrm{m}}}{27 f_{\mathrm{v} 0} \bar{H}_0} \ln \left(1+\frac{9 f_{\mathrm{v} 0} \bar{H}_0 t}{4 \Omega_{\mathrm{m}}}\right).
\end{equation}
We note that an average expansion law only holds on scales greater than the statistical homogeneity scale, which corresponds to a CMB rest frame redshift of order $z \sim 0.021-0.040$ \citep{Scrimgeour_2012, Ntelis_2017}. In this work we adopt the value used to quote the key results in \cite{Dam_2017} of $z_{\mathrm{min}} = 0.033$. We re-run the entire pipeline with this cut, which reduces our low-$z$ sample by 68 SNe (see Section~\ref{sec:data}; from here on we will refer to this modified sample as DES-SN5YR$_{\mathrm{cut}}$). We also use CMB rest-frame redshifts excluding peculiar velocity corrections of the host galaxy, which are calculated assuming a standard FLRW model to remain consistent with previous work by \cite{Dam_2017}. 

Finally, we retest the Flat-$\Lambda$CDM model with these same changes to make a consistent comparison between the two models. In addition to the above changes to the DES-SN5YR data, we also note that the conversion of redshift increments to a radial comoving distance involves different assumptions about spatial curvature in the FLRW and Timescape models (see Appendix D2 from \cite{Dam_2017} for more details). Therefore, we do not include the \cmb summary statistic as outlined in Section~\ref{cmbprior} when constraining the Timescape and Flat-$\Lambda$CDM models and include only angular measurements on the BAO scale (\baoperp from here on) from the SDSS data ($D_M(z_*)/D_M(z)$ constraints from Table~\ref{tab:BAOdata}).

Using DES-SN5YR alone, we find $f_{v0}=0.791^{+0.039}_{-0.034}$, equivalent to a dressed matter density of $\Omega_{\mathrm{m}} = 0.292^{+0.043}_{-0.051}$ and for Flat-$\Lambda$CDM find $\Omega_{\mathrm{m}} = 0.362^{+0.019}_{-0.018}$. These results are consistent with constraints found by \cite{Dam_2017} using the JLA catalogue \citep{2014JLA}. Fig.~\ref{fig:om_comparison} shows consistent matter density predictions between Flat-$\Lambda$CDM in the baseline analysis and Flat-$\Lambda$CDM$_{\mathrm{cut}}$ after including a redshift cut at $z_{\rm min}=0.033$ and excluding peculiar velocity corrections. However both are just outside the 68 per cent confidence interval of the Planck TTTEEE-lowE prediction \citep[$\Omega_{\mathrm{m}} =0.3166\pm 0.0084$;][]{2020_planck}. In contrast, the Timescape model has a lower central value for the matter density in agreement with Planck.

\begin{figure}\centering
    \includegraphics[width=0.8\linewidth]{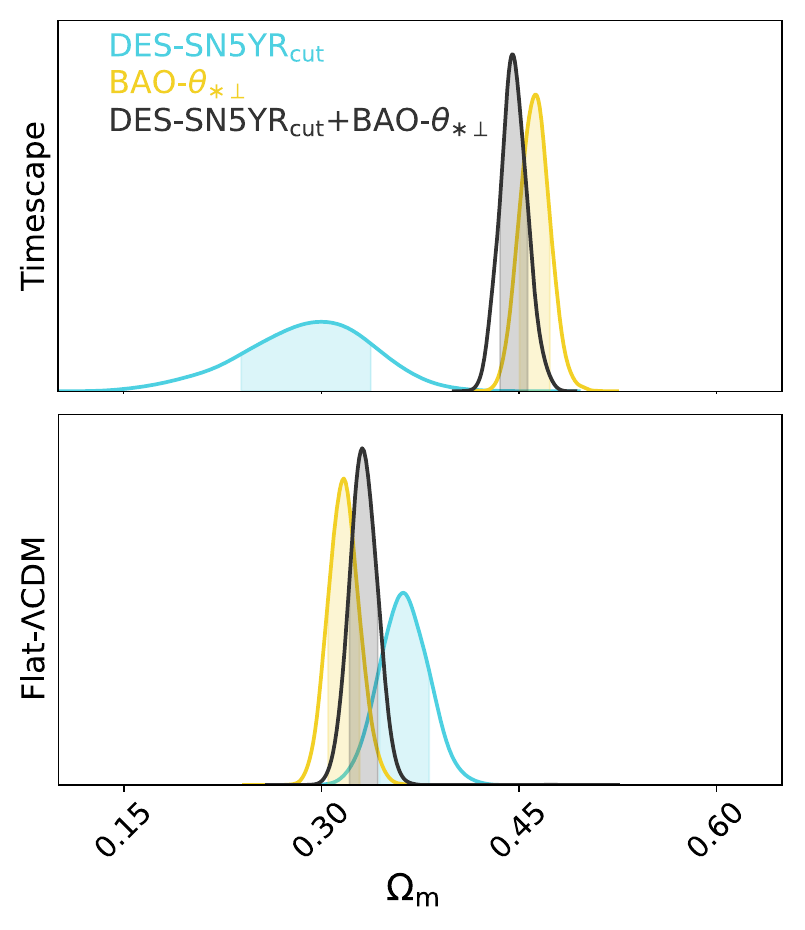}
    \caption{Constraints on the matter density from the DES-SN5YR$_{\rm cut}$ data set only (blue) and \baoperp (yellow) as well as the DES-SN5YR$_{\rm cut}$ combined with \baoperp prior (black). We show both the constraints from the Timescape and Flat-$\Lambda$CDM models (Section~\ref{sec:timescape}) with the same modifications to the data. In particular, we apply a redshift cut of $z_{\mathrm{min}}=0.033$ and excluding peculiar velocity corrections. Note that for Timescape cosmology, the void fraction is related to the dressed matter density by $\Omega_{\mathrm{m}}=\frac{1}{2}\left(1-f_{\mathrm{v} 0}\right)\left(2+f_{\mathrm{v} 0}\right)$. }
    \label{fig:contour_TSFLCDM}
\end{figure}

When combining the DES-SN5YR with \baoperp, we find $\Omega_{\mathrm{m}} = 0.446^{+0.010}_{-0.009}$ for the Timescape model and for Flat-$\Lambda$CDM find $\Omega_{\mathrm{m}} = 0.332^{+0.011}_{-0.010}$. These results are shown in Fig.~\ref{fig:contour_TSFLCDM}. It is apparent from the upper panel that the datasets \baoperp and DES-SN5YR are in tension in the Timescape model, and this model is therefore disfavoured relative to Flat-$\Lambda$CDM by the AIC statistic.

\section{Discussion}\label{sec:discussion}

\begin{table*}
\caption{Goodness of fit \& Model Comparison statistics. A more negative $\frac{1}{2}\Delta$AIC and $\Delta {\rm ln}~S$ value indicates a stronger preference over Flat-$\Lambda$CDM.}
\label{tab:model_comp}
\renewcommand{\arraystretch}{1.3}
\begin{tabular}{cccccccc}
\hline \hline
& \multicolumn{3}{c}{\textbf{DES-SN5YR}} & & \multicolumn{3}{c}{\textbf{DES-SN5YR~+~\cmb+~\bao}} \\[-1mm]

Model &  $\frac{1}{2}\Delta$AIC & $\Delta {\rm ln}~S$ & $\chi^2 $ & Model &  $\frac{1}{2}\Delta$AIC & $\Delta {\rm ln}~S$ & $\chi^2$  \\ [-1mm]\hline
        Cosmography - Third Order  & $-0.9$ &  $-1.37$  & 1641 &   &  &     &  \\
        Cosmography - Fourth Order  & $-3.6$ &  $-4.39$  & 1633 &   &  &     &  \\
        Flat-$\Lambda$CDM  & $0.0$ &  $0.0$  & 1645 & Flat-$\Lambda$CDM  & $0.0$ &  $0.0$   & 1665 \\
        $\Lambda$CDM & $0.6$ &  $0.09$  & 1644 &  $\Lambda$CDM & $0.4$ &  $-0.10$  & 1664 \\
        $w$CDM  & $1.1$ &  $0.13$    &   1643  &$w$CDM  & $-3.1$ &  $-3.64$    &   1655 \\
        Flat-$w_0 w_z$CDM  & $-1.8$ &  $-2.97$  & 1637 &Flat-$w_0 w_z$CDM  & $-3.1$ &  $-4.16$    &   1655  \\
        Flat-$w^{p}_{a}$CDM  & $-1.8$ &  $-2.58$ & 1637  &Flat-$w^{p}_{a}$CDM  & $-3.2$ & $-4.17$   &   1655 \\
        Thaw  & $1.0$ & $-0.57$   & 1643 & Thaw  & $-3.2$ &  $-4.60$  &   1655  \\
        SCG  & $0.9$ & $0.35$   & 1644  & SCG  & $138.4$ &  $138.03$  &   1940  \\
        FGCG  & $0.4$ &  $-0.30$  & 1643  & FGCG  & $-3.4$ & $-3.94$  &   1657 \\
        GCG  & $0.4$ &  $0.08$ & 1641   & GCG  & $-2.7$ &  $-3.71$  &  1656 \\
        NGCG  & $0.6$ &  $0.03$ & 1642 & NGCG  & $-3.2$ & $-4.08$ &   1655\\
        MPC  & $-1.8$ &  $-2.49$ & 1637  & MPC  & $-3.2$ &  $-3.94$  & 1655 \\
        IDE1  & $1.3$ &  $-0.17$ & 1643  & IDE1  & $-2.7$ &  $-3.70$  & 1656  \\
        IDE2  & $0.7$ &  $-0.23$ & 1642  & IDE2  & $-2.7$ & $-3.75$  & 1656  \\
        IDE3  & $0.1$ &  $-0.26$  & 1641  & IDE3  & $-3.2$ & $-3.82$  & 1655  \\
        DGP  & $0.6$ &  $-0.05$  & 1644  &DGP  & $31.5$ &  $31.11$ & 1726  \\
        GAL  & $0.9$ &  $0.34$  & 1644   &GAL  & $72.5$ & $72.10$  & 1808  \\ \hline
        & \multicolumn{3}{c}{\textbf{DES-SN5YR$_{\mathrm{cut}}$}} & & \multicolumn{3}{c}{\textbf{DES-SN5YR$_{\mathrm{cut}}$~+~\baoperp}} \\[-1mm] \hline
        Flat-$\Lambda$CDM  & $0.0$ &  $0.0$  &  1616 &Flat-$\Lambda$CDM  & $0.0$ &  $0.0$  & 1624 \\
        Timescape  & $-1.7$ &  $-1.72$  & 1612  &Timescape  & $6.3$ &  $6.17$  & 1637  \\
\hline
\end{tabular}
\end{table*}

\subsection{Goodness of fit}\label{sec:goodness}
To investigate the goodness of fit for each of the models we present the $\chi^2$ for various data combinations, see Table~\ref{tab:model_comp}, where $\chi^2=-2 \ln \mathcal{L}^{\max }$ and $\mathcal{L}^{\rm max}$ is the maximum likelihood of the entire parameter space. 

The number of degrees of freedom ($N_{\rm dof}$) is equal to the number of data points minus the number of cosmological parameters constrained for each model. For DES-SN5YR and DES-SN5YR$_{\mathrm{cut}}$, we approximate the number of data points by summing the BEAMS probability of each SN being Type Ia and find $\sum P_{\mathrm{B}(\mathrm{Ia})}=1735$ and 1666 respectively. The additional number of data points when including the \cmb, \bao or \baoperp are 1, 14 and 7 respectively. 

Using DES-SN5YR alone, we find that all models tested within this work result in good fits to the data. However, the SCG, DGP and GAL models have a poor $\chi^2$ when combining DES-SN5YR with the \cmb and \bao as they are unable to reconcile the additional data sets. To a lesser extent this also afflicts the Timescape model. This can be seen visually in Figs.~\ref{fig:contour_CG}a,~\ref{fig:contour_MPCIDE}e,~\ref{fig:contour_MPCIDE}f,~\&~\ref{fig:contour_TSFLCDM} where the parameter space of the combined contours do not share a common region with all probes.

\subsection{Model comparisons}\label{sec:model_comp}
To assess whether additional parameters invoked in the more complex models are justified given the data, we use the Akaike Information Criterion ${\rm AIC}\equiv 2k-2\ln\mathcal{L}^{\max }$ \citep{1100705}, where $k$ is the number of parameters in the model. We also use the Suspiciousness \citep{handley19}, which is defined as $\ln S= \ln R - \ln I$ where $R$ is the Bayes Ratio and $I$ is the Bayesian information. \cite{handley19} note that the Bayes ratio is prior-dependent and show that Suspiciousness is prior-independent due to the combination with the Bayesian information.

In Table~\ref{tab:model_comp} we quote the $\frac{1}{2}\Delta$AIC\footnote{We quote $\frac{1}{2}\Delta$AIC result, which allows us to use the same scale as the Suspiciousness.} and the difference in the logarithm of the Suspiciousness, $\Delta {\rm ln}~S$ relative to Flat-$\Lambda$CDM. To asses the strength of this preference, \citet{Trotta08} suggests that $\Delta>1$, $\Delta>2.5$ and $\Delta>5$ indicates weak, moderate and strong evidence respectively, against the model with the higher $\Delta$ value. In both cases, more negative values indicate that the data prefers the extended model over Flat-$\Lambda$CDM. We determine ln$~S$ using {\sc anesthetic} software \citep{anesthetic} with the nested sampling outputs from {\sc polychord} \citep{Handley15a, Handley15b} with $25\times k$ live points, $5\times k$ repeats and an evidence tolerance requirement of $0.1$. 

Using the DES-SN5YR alone, both the AIC and Suspiciousness find no strong evidence for or against any of the non-standard models. Both find weak evidence for the third order cosmographic model and moderate evidence for the fourth order cosmographic model. Furthermore, the AIC and Suspiciousness weakly and moderately prefer the Flat-$w_0 w_z$CDM, Flat-$w^{p}_{a}$CDM and MPC models over Flat-$\Lambda$CDM respectively. The Timescape model, which was fit using the DES-SN5YR$_{\rm cut}$ sample is weakly preferred by both the AIC and Suspiciousness.

When combined with the \cmb and \bao both the AIC and Suspiciousness agree that 11 of the 15 non-standard models we investigate are moderately preferred over Flat-$\Lambda$CDM. We note that this is not a result of curvature alone with no preference for or against the $\Lambda$CDM model. The top performing models include Flat-$w^{p}_{a}$CDM with $(\frac{1}{2}\Delta\mathrm{AIC}, \Delta{\rm ln}~S)=(-3.2, -4.17)$ indicating an evolution of $w$ that increases with time, the thawing model with $(\frac{1}{2}\Delta\mathrm{AIC}, \Delta{\rm ln}~S)=(-3.2, -4.60)$ and the FGCG model with $(\frac{1}{2}\Delta\mathrm{AIC}, \Delta{\rm ln}~S)=(-3.4, -3.94)$, which invokes an exotic background fluid. These results suggest that additional flexibility in our cosmological models may be required beyond the cosmological constant. 

\subsection{Tension metrics}\label{sec:suspiciousness}
We also use the Suspiciousness to assess whether different datasets are consistent (in contrast to Section~\ref{sec:model_comp} and Table~\ref{tab:model_comp} where the Suspiciousness was used as a model comparison statistic), which is ideal for cases such as ours where we have chosen deliberately wide and uninformative priors \citep[See][section 4.2]{lemos21}. We use the {\sc anesthetic} software \citep{handley19} to determine ln$~S$ and produce and ensemble of realisations to estimate sample variance. Using the scale from \citet{Trotta08}, ln$~S < -5$ is considered strong tension, $-5 < $ ln$~S < -2.5$ is considered moderate tension and ln$~S > -2.5$ indicates that the datasets are in agreement.
 \begin{figure}
    \centering
    \includegraphics[width=\linewidth]{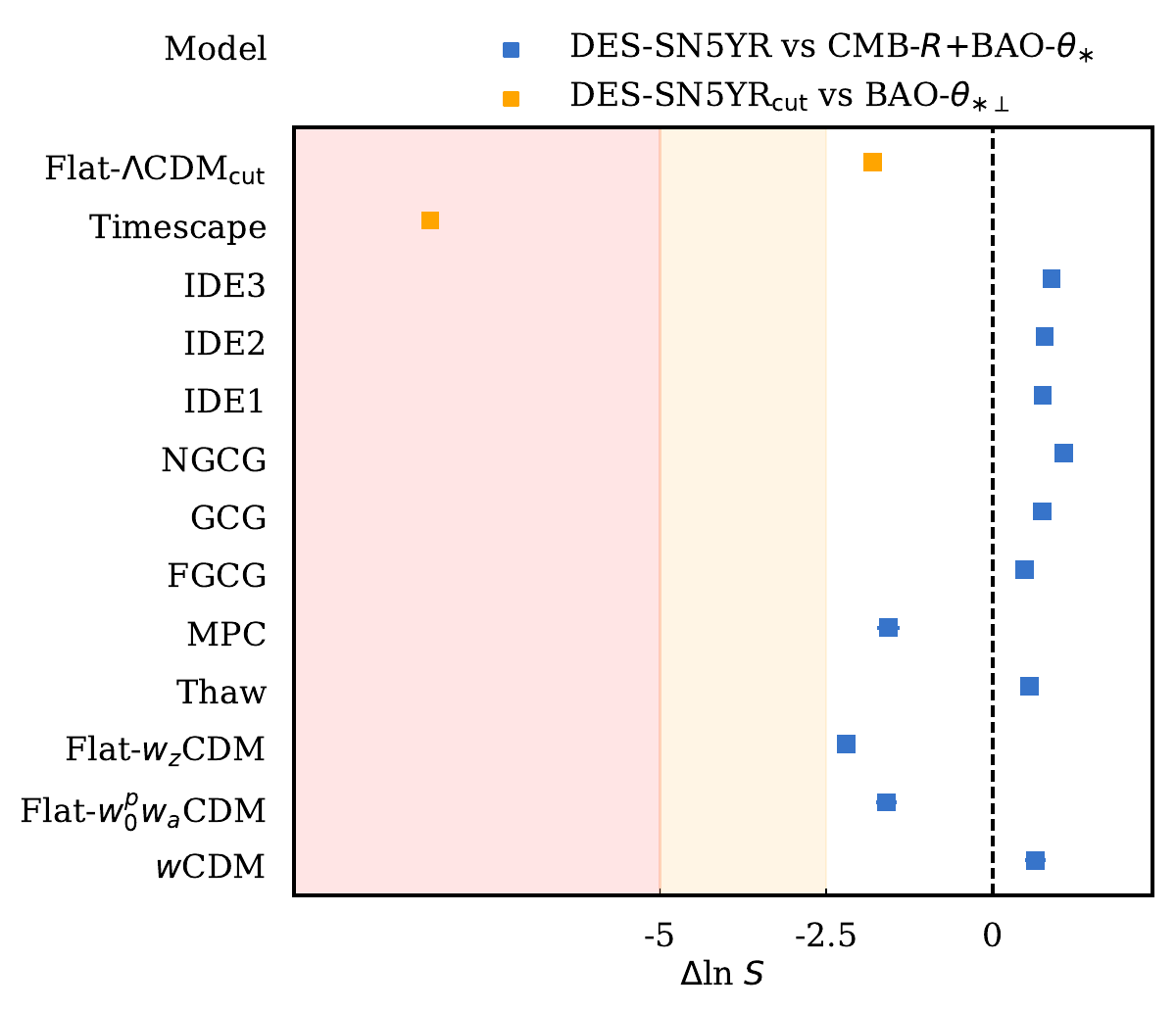}
    \caption{Measurements of $\Delta$ln$~S$ between the DES-SN5YR and the combined \cmb+~\bao datasets (blue). The modified data sets for the Timescape and Flat-$\Lambda$CDM$_{\mathrm{cut}}$ are shown in orange. The shaded yellow and red regions represent moderate and strong tension respectively. Note, models already been shown in Section~\ref{sec:goodness} to be poor fits to the combined data sets (SCG, DGP and GAL) have been excluded from the plot for clarity and all had $\Delta$ln$~S << -5$.}
    \label{fig:tensions}
\end{figure}
In Fig.~\ref{fig:tensions}, we plot the $\Delta$ln$~S$ between the relevant datasets. Note, models already been shown in Section~\ref{sec:goodness} to be poor fits to the combined data sets (SCG, DGP and GAL) have been excluded from the plot and all had $\Delta$ln$~S << -5$. We find a strong tension between the DES-SN5YR with \baoperp datasets when fitting the Timescape model. For all other models, we find no indication of tension.
 \begin{figure}
    \centering
    \includegraphics[width=\linewidth]{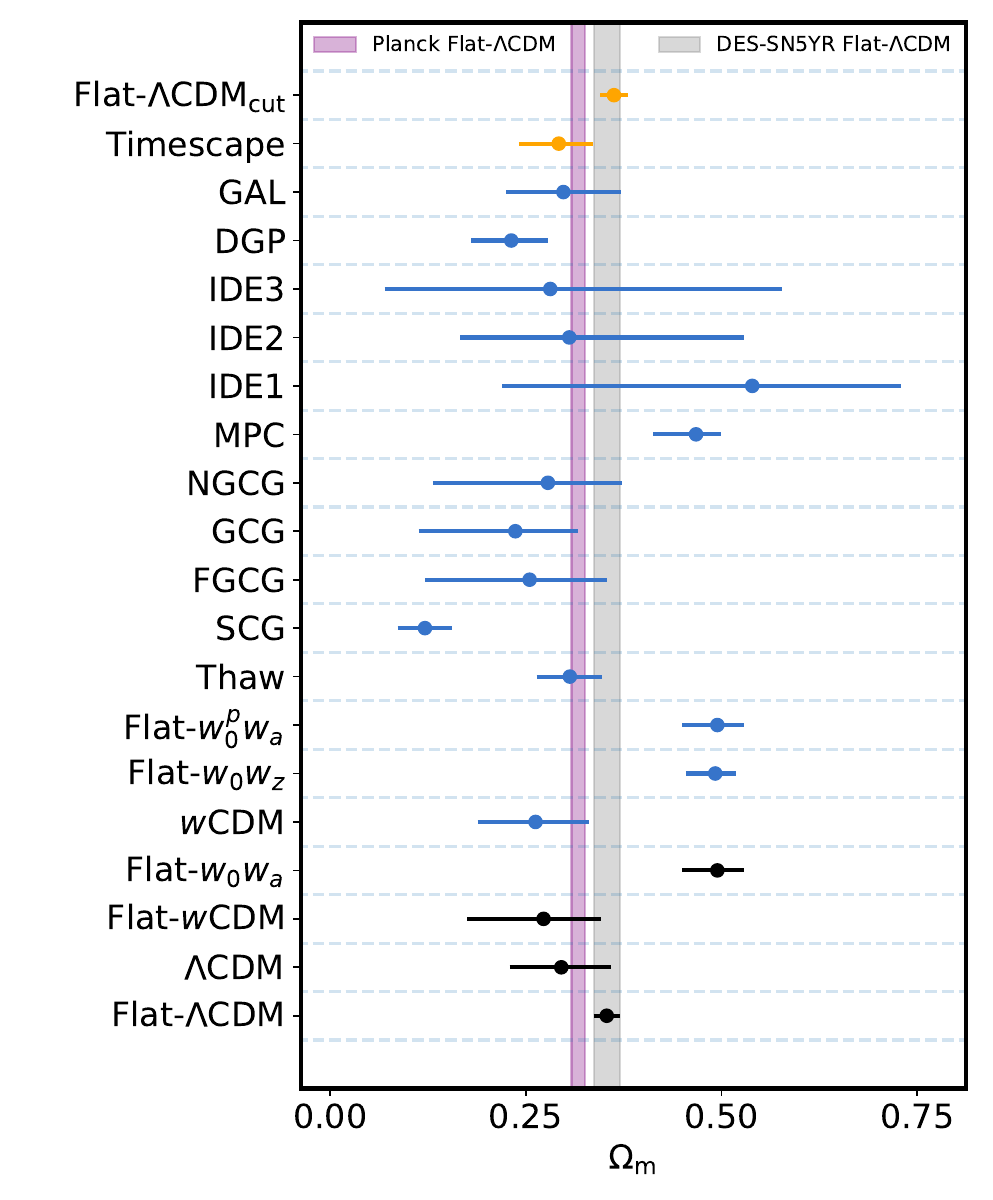}
    \caption{A summary of the best fit matter density for the models constrained by the DES-SN5YR sample. In black are the constraints from \KP, in blue and orange are constraints from this work, where the orange points highlight that the Hubble diagram used to constrain the Timescape and Flat-$\Lambda$CDM$_{\mathrm{cut}}$ models included a redshift cut at $z_{\mathrm{min}}=0.033$ and excluded peculiar velocity corrections. The purple shaded region represents the TTTEEE-lowE 68 per cent confidence limits for the Flat-$\Lambda$CDM model determined by the Planck collaboration \citep{2020_planck}.}
    \label{fig:om_comparison}
\end{figure}

\section{Conclusions}\label{sec:conclusions}
The DES Supernova survey is the largest, most homogeneous SN data set to date containing 1635 supernovae combined with 194 existing Low-$z$ SNe Ia. The statistical power of the DES-SN5YR sample allows us to obtain robust and precise constraints on cosmological models beyond $\Lambda$CDM.

We first investigated two important areas of the main DES supernova cosmology analysis that are, or may appear to be subject to cosmological dependencies. 

\begin{enumerate}
    \item We demonstrated that the assumption of a reference cosmology used to generate simulated light curves and perform selection bias corrections to the data results in a bias that is sub-dominant to statistical uncertainties. 
    For non-standard models, we also show a region of expansion histories where we are confident in our bias corrections. For the next era of SN experiments, the reference cosmology may become a dominating systematic and as a result, we show that an iterative method (where the reference cosmology is updated in a second iteration based on the best fit cosmology from the first) is viable and can be employed to reduce this bias. 
    
    \item We demonstrated that the BBC fitting procedure, which uses a fiducial cosmology, is insensitive to that choice of cosmology. We also show that a spline is viable alternative to a fiducial cosmology as it obtains consistent results and may reduce confusion as to the role of the fiducial cosmology in future analyses. 
\end{enumerate}

Secondly, we presented constraints on 15 exotic cosmological models using the DES-SN5YR sample alone and after combining the DES-SN5YR with external probes. Using DES-SN5YR alone, we find that all models tested within this work are good fits to the data. This trend continues when we combine the DES-SN5YR with priors from the \cmb and \bao except for models that had been previously ruled out. We assessed whether additional parameters invoked in the more complex models are justified given the data by using the Akaike Information Criteria and Suspiciousness. Of the 15 models that we test, we find no strong evidence for or against any of the non-standard models for any of our data combinations. Using the DES-SN5YR alone, the Suspiscousness moderately prefers 3 of the non-standard models along with the fourth order cosmographic model. When combined with the \cmb and \bao both the AIC and Suspiciousness agree that 11 models are moderately preferred over Flat-$\Lambda$CDM. We show that this is not a result of curvature alone. Our work suggests that additional flexibility in our cosmological models may be required beyond the cosmological constant.

\section*{Acknowledgements}

{\footnotesize 
\textbf{Author Contributions:} RCa developed the code, performed the analysis, drafted the manuscript. TMD conceived the idea, guided the project, ran test code and helped writing. MV contributed to developing part of the analysis, guided the project and commented on drafts. PS commented on drafts, ran test code and helped writing. All authors contributed to this paper and/or carried out infrastructure work that made this analysis possible. Highlighted contributions include: 
\textit{Contributed to project development:} 
DB, DS, JF, RK. \textit{Contributed to advising and validation:} 
CL, DS, JF, MSa, PA, PW, RK. \textit{Construction and validation of the DES-SN5YR Hubble diagram:} 
AC, AM, BP, BS, CL, DB, DS, GT, HQ, JL, LG, MSa, MSm, MSu, MT, PA, PW, RCh, RK SH. \textit{Contributed to the internal review process:} 
CL, GFL, JL, LG, MSu, PW, RK. The remaining authors have made contributions to this paper that include, but are not limited to, the construction of DECam and other aspects of collecting the data; data processing and calibration; developing broadly used methods, codes, and simulations; running the pipelines and validation tests; and promoting the science analysis.

TMD, AC, RCa, SH, acknowledge the support of an Australian Research Council Australian Laureate Fellowship (FL180100168) funded by the Australian Government. JF would like to acknowledge Shrihan Agarwal and Yuuki Omori for helpful conversations. MV was partly supported by NASA through the NASA Hubble Fellowship grant HST-HF2-51546.001-A awarded by the Space Telescope Science Institute, which is operated by the Association of Universities for Research in Astronomy, Incorporated, under NASA contract NAS5-26555. LG acknowledges financial support from the Spanish Ministerio de Ciencia e Innovaci\'on (MCIN) and the Agencia Estatal de Investigaci\'on (AEI) 10.13039/501100011033 under the PID2020-115253GA-I00 HOSTFLOWS project, from Centro Superior de Investigaciones Cient\'ificas (CSIC) under the PIE project 20215AT016 and the program Unidad de Excelencia Mar\'ia de Maeztu CEX2020-001058-M, and from the Departament de Recerca i Universitats de la Generalitat de Catalunya through the 2021-SGR-01270 grant. AM is supported by the ARC Discovery Early Career Researcher Award (DECRA) project number DE230100055.

Funding for the DES Projects has been provided by the U.S. Department of Energy, the U.S. National Science Foundation, the Ministry of Science and Education of Spain, 
the Science and Technology Facilities Council of the United Kingdom, the Higher Education Funding Council for England, the National Center for Supercomputing 
Applications at the University of Illinois at Urbana-Champaign, the Kavli Institute of Cosmological Physics at the University of Chicago, 
the Center for Cosmology and Astro-Particle Physics at the Ohio State University,
the Mitchell Institute for Fundamental Physics and Astronomy at Texas A\&M University, Financiadora de Estudos e Projetos, 
Funda{\c c}{\~a}o Carlos Chagas Filho de Amparo {\`a} Pesquisa do Estado do Rio de Janeiro, Conselho Nacional de Desenvolvimento Cient{\'i}fico e Tecnol{\'o}gico and 
the Minist{\'e}rio da Ci{\^e}ncia, Tecnologia e Inova{\c c}{\~a}o, the Deutsche Forschungsgemeinschaft and the Collaborating Institutions in the Dark Energy Survey. 

The Collaborating Institutions are Argonne National Laboratory, the University of California at Santa Cruz, the University of Cambridge, Centro de Investigaciones Energ{\'e}ticas, 
Medioambientales y Tecnol{\'o}gicas-Madrid, the University of Chicago, University College London, the DES-Brazil Consortium, the University of Edinburgh, 
the Eidgen{\"o}ssische Technische Hochschule (ETH) Z{\"u}rich, 
Fermi National Accelerator Laboratory, the University of Illinois at Urbana-Champaign, the Institut de Ci{\`e}ncies de l'Espai (IEEC/CSIC), 
the Institut de F{\'i}sica d'Altes Energies, Lawrence Berkeley National Laboratory, the Ludwig-Maximilians Universit{\"a}t M{\"u}nchen and the associated Excellence Cluster Universe, 
the University of Michigan, NSF's NOIRLab, the University of Nottingham, The Ohio State University, the University of Pennsylvania, the University of Portsmouth, 
SLAC National Accelerator Laboratory, Stanford University, the University of Sussex, Texas A\&M University, and the OzDES Membership Consortium.

Based in part on observations at Cerro Tololo Inter-American Observatory at NSF's NOIRLab (NOIRLab Prop. ID 2012B-0001; PI: J. Frieman), which is managed by the Association of Universities for Research in Astronomy (AURA) under a cooperative agreement with the National Science Foundation.

The DES data management system is supported by the National Science Foundation under Grant Numbers AST-1138766 and AST-1536171.
The DES participants from Spanish institutions are partially supported by MICINN under grants ESP2017-89838, PGC2018-094773, PGC2018-102021, SEV-2016-0588, SEV-2016-0597, and MDM-2015-0509, some of which include ERDF funds from the European Union. IFAE is partially funded by the CERCA program of the Generalitat de Catalunya.
Research leading to these results has received funding from the European Research
Council under the European Union's Seventh Framework Program (FP7/2007-2013) including ERC grant agreements 240672, 291329, and 306478.
We  acknowledge support from the Brazilian Instituto Nacional de Ci\^encia
e Tecnologia (INCT) do e-Universo (CNPq grant 465376/2014-2).

This manuscript has been authored by Fermi Research Alliance, LLC under Contract No. DE-AC02-07CH11359 with the U.S. Department of Energy, Office of Science, Office of High Energy Physics.
}

\textit{Software:}
{\sc numpy} \citep{numpy}, 
{\sc astropy} \citep{astropy13,astropy18}, 
{\sc uncertainties} \citep{uncertainties},
{\sc matplotlib} \citep{matplotlib}, 
{\sc pandas} \citep{pandas}, 
{\sc scipy} \citep{scipy}, 
{\sc snana} \citep{kessler09}, 
{\sc pippin} \citep{Hinton2020}, 
{\sc chainconsumer} \citep{hinton16},
{\sc cobaya} \citep{cobaya2, cobaya1},
{\sc polychord} \citep{Handley15a, Handley15b},
{\sc anesthetic} \citep{anesthetic}.

\section*{Data Availability}

The MCMC and {\sc polychord} chains along with scripts to reproduce our results in Tables~\ref{tab:results}~\&~\ref{tab:model_comp} and contour plots can be found at \url{https://github.com/RyanCamo/DESSN_extensions/}.


\bibliographystyle{mnras}
\bibliography{mybib} 


\noindent \\ $^{1}$ School of Mathematics and Physics, University of Queensland,  Brisbane, QLD 4072, Australia\\
$^{2}$ Department of Physics, Duke University Durham, NC 27708, USA\\
$^{3}$ Department of Physics \& Astronomy, University College London, Gower Street, London, WC1E 6BT, UK\\
$^{4}$ Fermi National Accelerator Laboratory, P. O. Box 500, Batavia, IL 60510, USA\\
$^{5}$ Kavli Institute for Cosmological Physics, University of Chicago, Chicago, IL 60637, USA\\
$^{6}$ Department of Astronomy and Astrophysics, University of Chicago, Chicago, IL 60637, USA\\
$^{7}$ The Research School of Astronomy and Astrophysics, Australian National University, ACT 2601, Australia\\
$^{8}$ Center for Astrophysics $\vert$ Harvard \& Smithsonian, 60 Garden Street, Cambridge, MA 02138, USA\\
$^{9}$ Institute of Space Sciences (ICE, CSIC),  Campus UAB, Carrer de Can Magrans, s/n,  08193 Barcelona, Spain\\
$^{10}$ Centre for Astrophysics \& Supercomputing, Swinburne University of Technology, Victoria 3122, Australia\\
$^{11}$ Department of Physics and Astronomy, University of Pennsylvania, Philadelphia, PA 19104, USA\\
$^{12}$ Centre for Gravitational Astrophysics, College of Science, The Australian National University, ACT 2601, Australia\\
$^{13}$ School of Physics and Astronomy, University of Southampton,  Southampton, SO17 1BJ, UK\\
$^{14}$ Aix Marseille Univ, CNRS/IN2P3, CPPM, Marseille, France\\
$^{15}$ Cerro Tololo Inter-American Observatory, NSF's National Optical-Infrared Astronomy Research Laboratory, Casilla 603, La Serena, Chile\\
$^{16}$ Laborat\'orio Interinstitucional de e-Astronomia - LIneA, Rua Gal. Jos\'e Cristino 77, Rio de Janeiro, RJ - 20921-400, Brazil\\
$^{17}$ Department of Physics, University of Michigan, Ann Arbor, MI 48109, USA\\
$^{18}$ Institut de F\'{\i}sica d'Altes Energies (IFAE), The Barcelona Institute of Science and Technology, Campus UAB, 08193 Bellaterra (Barcelona) Spain\\
$^{19}$ Institute of Cosmology and Gravitation, University of Portsmouth, Portsmouth, PO1 3FX, UK\\
$^{20}$ CNRS, UMR 7095, Institut d'Astrophysique de Paris, F-75014, Paris, France\\
$^{21}$ Sorbonne Universit\'es, UPMC Univ Paris 06, UMR 7095, Institut d'Astrophysique de Paris, F-75014, Paris, France\\
$^{22}$ University Observatory, Faculty of Physics, Ludwig-Maximilians-Universit\"at, Scheinerstr. 1, 81679 Munich, Germany\\
$^{23}$ Kavli Institute for Particle Astrophysics \& Cosmology, P. O. Box 2450, Stanford University, Stanford, CA 94305, USA\\
$^{24}$ SLAC National Accelerator Laboratory, Menlo Park, CA 94025, USA\\
$^{25}$ Instituto de Astrofisica de Canarias, E-38205 La Laguna, Tenerife, Spain\\
$^{26}$ Institut d'Estudis Espacials de Catalunya (IEEC), 08034 Barcelona, Spain\\
$^{27}$ Hamburger Sternwarte, Universit\"{a}t Hamburg, Gojenbergsweg 112, 21029 Hamburg, Germany\\
$^{28}$ Department of Physics, IIT Hyderabad, Kandi, Telangana 502285, India\\
$^{29}$ Jet Propulsion Laboratory, California Institute of Technology, 4800 Oak Grove Dr., Pasadena, CA 91109, USA\\
$^{30}$ Institute of Theoretical Astrophysics, University of Oslo. P.O. Box 1029 Blindern, NO-0315 Oslo, Norway\\
$^{31}$ Instituto de Fisica Teorica UAM/CSIC, Universidad Autonoma de Madrid, 28049 Madrid, Spain\\
$^{32}$ Santa Cruz Institute for Particle Physics, Santa Cruz, CA 95064, USA\\
$^{33}$ Center for Cosmology and Astro-Particle Physics, The Ohio State University, Columbus, OH 43210, USA\\
$^{34}$ Department of Physics, The Ohio State University, Columbus, OH 43210, USA\\
$^{35}$ Australian Astronomical Optics, Macquarie University, North Ryde, NSW 2113, Australia\\
$^{36}$ Lowell Observatory, 1400 Mars Hill Rd, Flagstaff, AZ 86001, USA\\
$^{37}$ Sydney Institute for Astronomy, School of Physics, A28, The University of Sydney, NSW 2006, Australia\\
$^{38}$ George P. and Cynthia Woods Mitchell Institute for Fundamental Physics and Astronomy, and Department of Physics and Astronomy, Texas A\&M University, College Station, TX 77843,  USA\\
$^{39}$ LPSC Grenoble - 53, Avenue des Martyrs 38026 Grenoble, France\\
$^{40}$ Instituci\'o Catalana de Recerca i Estudis Avan\c{c}ats, E-08010 Barcelona, Spain\\
$^{41}$ Perimeter Institute for Theoretical Physics, 31 Caroline St. North, Waterloo, ON N2L 2Y5, Canada\\
$^{42}$ Department of Astrophysical Sciences, Princeton University, Peyton Hall, Princeton, NJ 08544, USA\\
$^{43}$ Observat\'orio Nacional, Rua Gal. Jos\'e Cristino 77, Rio de Janeiro, RJ - 20921-400, Brazil\\
$^{44}$ Ruhr University Bochum, Faculty of Physics and Astronomy, Astronomical Institute, German Centre for Cosmological Lensing, 44780 Bochum, Germany\\
$^{45}$ Centro de Investigaciones Energ\'eticas, Medioambientales y Tecnol\'ogicas (CIEMAT), Madrid, Spain\\
$^{46}$ Computer Science and Mathematics Division, Oak Ridge National Laboratory, Oak Ridge, TN 37831\\
$^{47}$ Center for Astrophysical Surveys, National Center for Supercomputing Applications, 1205 West Clark St., Urbana, IL 61801, USA\\
$^{48}$ Department of Astronomy, University of California, Berkeley,  501 Campbell Hall, Berkeley, CA 94720, USA\\
$^{49}$ Lawrence Berkeley National Laboratory, 1 Cyclotron Road, Berkeley, CA 94720, USA\\

\bsp	
\label{lastpage}

\end{document}